\documentclass[twocolumn]{autart}    

\usepackage{graphicx}          
\usepackage{cite}
\usepackage{amsmath,amssymb,amsfonts}
\usepackage{graphicx}
\let\classAND\AND
\let\AND\relax
\usepackage{algorithmic}

\let\AND\classAND
\AtBeginEnvironment{algorithmic}{\let\AND\algoAND}
\usepackage{algorithm}
\usepackage{hyperref}
\hypersetup{hidelinks=true}
\usepackage{textcomp}

\usepackage{caption}
\usepackage{subcaption}

\usepackage{algorithm}
\usepackage{algorithmic}

\usepackage[textsize=footnotesize,textwidth=1.5cm]{todonotes}

\usepackage{mathrsfs}

\newcommand{\Domain}{\mathcal{D}}
\newcommand{\NUGIS}{\textbf{NUGIS}}

\newcommand{\PWA}{\mathrm{PWA}}

\newcommand{\genpwaparamM}{A}
\newcommand{\genpwaparamV}{a}

\newcommand{\Mode}{I}
\newcommand{\state}{x}

\newcommand{\pState}{X}

\newcommand{\Vertex}{\mathcal{F}_0}

\newcommand{\prtition}{\mathcal{P}}

\newcommand{\qState}{Z}

\newcommand{\vMatrix}{E}
\newcommand{\vVec}{e}

\newcommand{\Rprtition}{\mathcal R}

\newcommand{\Pindex}{\Mode(\prtition)}

\newcommand{\Int}[1]{\mathrm{Int}\left(#1 \right)}
\newcommand{\Dom}[1]{\mathrm{Dom}\left(#1 \right)}
\newcommand{\conv}[1]{\mathrm{conv} \left(#1 \right)}

\newcommand{\R}{\mathbb{R}}

\begin{document}

\begin{frontmatter}

\title{SEROAISE: Advancing ROA Estimation for ReLU and PWA Dynamics through Estimating Certified Invariant Sets} 

\thanks[footnoteinfo]{This paper was not presented at any IFAC 
meeting. Corresponding author Hasan A. Poonawala. hasan.poonawala@uky.edu}

\author[USA]{Pouya Samanipour}\ead{psa254@uky.edu},    
\author[USA]{Hasan A. Poonawala}\ead{hasan.poonawala@uky.edu},               

\address[USA]{University of Kentucky, USA}  


\begin{abstract}                          
This paper presents a novel framework for constructing the Region of Attraction (RoA) for dynamics derived either from Piecewise Affine ($\PWA$) functions or from Neural Networks (NNs) with Rectified Linear Units (ReLU) activation function. This method, described as Sequential Estimation of RoA based on Invariant Set Estimation (SEROAISE), computes a Lyapunov-like $\PWA$ function over a certified $\PWA$ invariant set. While traditional approaches search for Lyapunov functions by enforcing Lyapunov conditions over pre-selected domains, this framework enforces Lyapunov-like conditions over a certified invariant subset obtained using the Iterative Invariant Set Estimator(IISE). Compared to the state-of-the-art, IISE provides systematically larger certified invariant sets. 
In order to find a larger invariant subset, the IISE utilizes a novel concept known as the Non-Uniform Growth of Invariant Set ($\NUGIS$). A number of examples illustrating the efficacy of the proposed methods are provided, including dynamical systems derived from learning algorithms. The implementation is publicly available at: \url{https://github.com/PouyaSamanipour/SEROAISE.git}.
\end{abstract}

\end{frontmatter}
\section{Introduction}
Stability analysis of dynamical systems, particularly for piecewise affine ($\PWA$) systems, plays a critical role in control theory and various engineering applications. A key aspect of stability analysis is the estimation of the region of attraction (RoA), which defines the set of initial conditions from which the system will converge to a specific equilibrium point. 
Robust performance against disturbances and faults is essential in dynamical systems. Hence, enlarging the RoA is necessary, particularly in safety-critical applications like robotics, aerospace, and automotive control, where resilience to disturbances is important~\cite{liu2024tool,zou2022analysing}.

Lyapunov functions serve as a fundamental tool in nonlinear stability analysis by providing a means to estimate the RoA. Recently, there has been significant interest in learning Lyapunov functions using neural networks (NNs), particularly with ReLU activation functions due to their $\PWA$ nature. Such approaches have been applied both to dynamics identified via NNs~\cite{dai2021lyapunov} and to NN-based controllers~\cite{dai2021lyapunov,yanglyapunov,wu2023neural}. Nonetheless, determining a suitable Lyapunov function for these ReLU-based dynamics remains challenging, primarily because of their inherent nonlinearity. This challenge is also prevalent in explicit model predictive control (MPC), where the control law is $\PWA$ and finding Lyapunov functions for stability analysis continues to be an open problem~\cite{bemporad2000piecewise,9992902}.
\begin{figure}[tbp]
    \centering
    \begin{subfigure}{0.5\textwidth}
        \centering
        \includegraphics[scale=0.42]{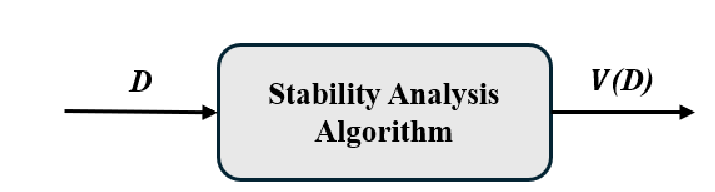}
        \caption{Most algorithms search for Lyapunov functions by enforcing Lyapunov conditions over a pre-selected domain $\Domain$~\cite{10108069,10313502}. }
        \label{fig:Whole dom}
    \end{subfigure}
    \hfill
    \begin{subfigure}{0.5\textwidth}
        \centering
        \includegraphics[scale=0.53]{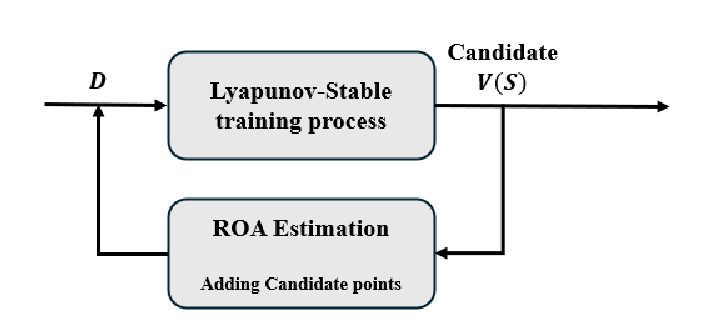}
        \caption{A novel framework in~\cite{yanglyapunov} proposes enforcing Lyapunov conditions on an invariant subset obtained by learning a candidate Lyapunov function.} 
        \label{fig:sampling based method}
    \end{subfigure}
    \hfill
    \begin{subfigure}{0.5\textwidth}
        \centering
        \includegraphics[scale=0.53]{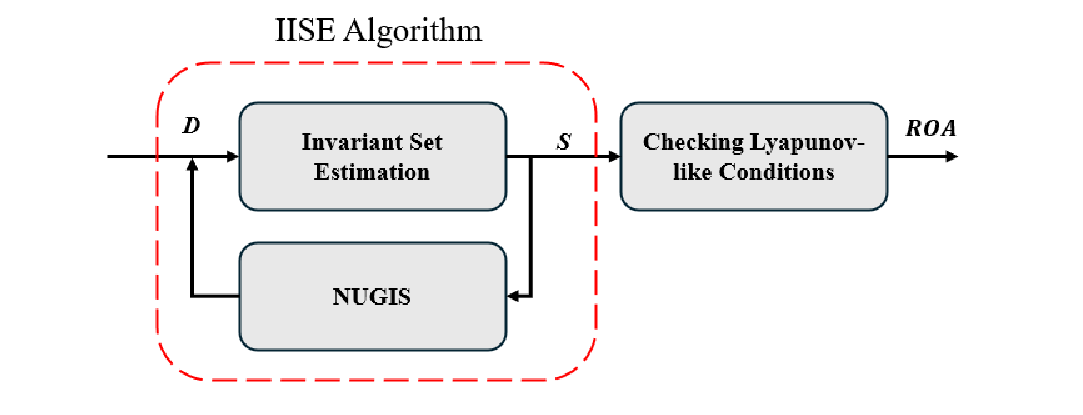}
        \caption{We propose SEROAISE, where an invariant subset $S$ of a pre-selected domain is identified by searching for a barrier function over that domain and then searching for RoA by enforcing Lyapunov-like conditions on only $S$.}
        \label{fig:proposed method}
    \end{subfigure}
    \caption{Three different approaches for selection of the domain  over which Lyapunov stability conditions are enforced when computationally searching for Lyapunov functions.}
    \label{fig:main}
\end{figure}
Several computational approaches have been developed to address these challenges. The Sum of Squares (SOS) method, implemented through semidefinite programming (SDP), provides a framework for analyzing local stability and estimating RoAs~\cite{iannelli2019robust}. Another common strategy employs piecewise quadratic (PWQ) Lyapunov functions~\cite{iervolino2020asymptotic,10108069}.
A critical limitation of these methods lies in the inherent conservatism introduced by the S-procedure. The conservatism can be mitigated by using piecewise affine ($\PWA$) Lyapunov functions, which can be formulated without additional conservatism, as opposed to their PWQ counterparts. Previously, we developed a linear optimization problem based on $\PWA$ parameterizations of the Lyapunov function over the pre-selected domain~\cite{10313502,10108069}. 

Beyond these structured function classes, neural networks (NNs) provide a flexible alternative for learning Lyapunov functions. According to the universal approximation theorem~\cite{gorban1998generalized}, NNs can approximate any continuous function arbitrarily well, making them well-suited for learning Lyapunov functions from finite samples. Sampling-based methods have been particularly successful in identifying Lyapunov functions that satisfy stability conditions in the state space~\cite{abate2020formal,korda2022stability,ravanbakhsh2017learning}. However, these approaches require verification to ensure correctness, which can be performed inexactly through relaxed convex formulations~\cite{fazlyab2020safety} or exactly using formal methods such as Satisfiability Modulo Theories (SMT) and Mixed-Integer Programming (MIP)~\cite{abate2020formal,chang2019neural,chen2021learning,dai2021lyapunov,edwards2023fossil,edwards2023general,zhou2022neural,dawson2022safe}. 
Exact verification is critical for confirming Lyapunov conditions or identifying counterexamples to refine candidate functions. Despite their success, scalability and efficiency remain persistent challenges, particularly for high-dimensional systems.

Despite advancements in both PWQ and PWA Lyapunov formulations, as well as neural network-based approaches, existing methods typically apply the search procedure to the entire domain, as illustrated in Fig.~\ref{fig:Whole dom}.
Due to the fact that there is no guarantee that all the initial conditions in the domain will converge to equilibrium, checking Lyapunov conditions over the entire domain is a computationally inefficient and unnecessary step~\cite{yanglyapunov}. To check the Lyapunov conditions over an invariant set for discrete-time dynamics,~\cite{yanglyapunov} proposed a sampling-based method, as shown in Fig.~\ref{fig:sampling based method}. An iterative process was used to develop a certifiable RoA in that study. It is important to note that there is no guarantee that continuing to iterate will result in a larger RoA (see Section~\ref{sec:problemstatement}). In addition, the certified invariant set is derived by checking the future states in discrete-time dynamics, which are easier to compute.

To address these challenges for continuous $\PWA$ dynamics, we propose a new method that restricts the Lyapunov search algorithm to a certified invariant set. As shown in Fig.~\ref{fig:proposed method}, we introduce 	\textbf{S}equential \textbf{E}stimation of \textbf{RoA} based on \textbf{I}nvariant \textbf{S}et 	\textbf{E}stimation (\textbf{SEROAISE}), a sequential method to find $\PWA$ RoA over a certified invariant set using a $\PWA$ barrier function. An invariant set estimation is a modification of the method described in~\cite{samanipour2024invariant}. As an additional feature, the Non-Uniform Growth of the Invariant Set ($\NUGIS$) is introduced in order to facilitate the growth of the invariant set. 
Our key contributions are as follows:
\begin{itemize}
    \item We propose a novel concept called Non-uniform Growth of Invariant Sets ($\NUGIS$), which enables the invariant set to expand non-uniformly, providing greater flexibility in constructing larger sets.
    \item We present a novel framework, Iterative Invariant Set Estimation (IISE), for identifying certified invariant sets for ReLU-based dynamical systems. This framework builds on our previous work~\cite{samanipour2024invariant} and allows the estimation of significantly larger certified invariant sets for $\PWA$ dynamical systems.
    \item We propose \textbf{SEROAISE}, a sequential method that identifies a larger RoA compared to state-of-the-art techniques, validated through non-trivial examples.
\end{itemize}
\section{SEROAISE Overview}
\label{sec:problemstatement}
We are interested in determining the \textbf{Region of Attraction (RoA)} for dynamical systems driven by Piecewise Affine ($\PWA$) functions as follows:
\begin{equation}\label{eq:pwa dynamic}
    \dot{x} = \PWA(x), \quad x \in \mathcal{D}
\end{equation}
where $\PWA$ is defined over a pre-selected domain $\mathcal{D} \subset \R^n$.

In our previous work~\cite{10313502,10108069}, we tackled the challenge of \textbf{automatically obtaining a certified Lyapunov function and RoA} for $\PWA$ dynamical systems and their equivalent ReLU neural networks, as shown in Figure~\ref{fig:Whole dom}. Although effective, this method checks Lyapunov conditions across the entire pre-selected domain $\Domain$. As observed in~\cite{yanglyapunov}, checking the Lyapunov conditions at states in $\Domain$ that lie outside the region of attraction results in unnecessary constraints on the parameters of the Lyapunov function. These additional constraints introduce conservatism in the form of a smaller region of attraction. To avoid these issues, Yang \emph{et. al.}~\cite{yanglyapunov} enforce Lyapunov conditions on states belonging to an estimate of the RoA. The authors resolve the chicken-and-egg problem by initializing an estimate of the RoA using linear systems techniques near the origin, and then using samples from outside its boundary to update the parameters of a candidate Lyapunov function, enabling a cyclic bootstrapping process. Once this samples-based training cycle is ended, a sound verification technique is used to find a valid region of attraction as a level set of the candidate Lyapunov function. 

Rather than relying on Lyapunov stability conditions to find the RoA, our objective is to relax these requirements to \emph{barrier function} conditions. 
The region of attraction of an equilibrium is also a forward invariant set. Instead of estimating this RoA using Lyapunov stability conditions, we may use conditions for a barrier function to find a certified invariant set.
This idea is motivated by LaSalle’s theorem, which ensures asymptotic stability in invariant sets.
\begin{thm}[LaSalle’s theorem~\cite{khalil2015nonlinear}]\label{th:Lasalle}
    Let $S \subseteq D \subset \mathbb{R}^n$ be a compact set that is forward invariant for the following dynamical systems.
    \begin{equation}\label{eq:general nl}
        \dot{x} = f_{cl}(x).
    \end{equation}
    Let $V : \mathcal{D} \to \mathbb{R}$ be a differentiable function such that $\dot{V}(x) \leq 0$ in $S$. Let $E \subseteq S$ be the set where $\dot{V}(x) = 0$. Let $\Omega \subseteq E$ be the largest invariant set in $E$. Then, every solution starting in $S$ approaches $\Omega$ as $t \to \infty$.
\end{thm}

\begin{cor}\label{cor:result lasalle}
    If a continuously differentiable function $V$ exists such that $\dot{V}(x) < 0$ in $S \setminus \{0\}$ and $\dot{V}(0) = 0$, then the \textbf{forward invariant set} $S$ estimates the RoA, i.e., $S \subseteq \text{RoA}$.
\end{cor}
\begin{pf}
    Let $x(t)$ be a solution of~\eqref{eq:general nl} starting in $S$. Given the assumption that $\dot{V}(x) < 0$  for all $x \in S \setminus \{0\}$ and $\dot{V}(0) = 0$, we conclude that $E = \{0\}$ , where $E$ is the set of points in $S$ at which $\dot{V}(x) = 0$. By Theorem~\ref{th:Lasalle}, every solution of~\eqref{eq:general nl} that starts within $S$ will approach the equilibrium point at the origin as $t \to \infty$. 
    Since $S$ is forward invariant, solutions starting in $S$ remain within $S$ for all future time. Thus, $S$ includes initial conditions that will converge to the equilibrium without leaving $S$. Therefore, $S \subseteq \text{RoA}$.
\end{pf}
\textbf{SEROAISE} is developed to take advantage of invariant set conditions in order to obtain larger RoA estimates.
\textbf{SEROAISE} introduces \textbf{a framework that finds a large RoA by identifying a large certified invariant subset and applying Lyapunov-like conditions to states inside this subset.} 
Our framework begins by finding a large certified invariant subset for $\PWA$ dynamic~\eqref{eq:pwa dynamic}. While~\cite{samanipour2024invariant} proposed a method for this, it has drawbacks:
\begin{enumerate}
    \item  \textbf{Guaranteed Growth:}\label{challenge:Gaurantee} If we aim to enlarge an invariant subset obtained using~\cite{samanipour2024invariant}, there is no guarantee that the new subset will exist.
    \item \textbf{Linear $\alpha(x)$:} \label{goal:linear alpha}
    Using the linear form of $\alpha(x)$ in barrier function constraints as described in~\cite{samanipour2024invariant} restricts the search space and excludes potentially larger invariant sets.
    \item \textbf{Boundary Point Exclusion:}\label{goal: BPI} The work in~\cite{samanipour2024invariant} searches for the invariant subset on the interior of $\Domain$, which is conservative compared to searching over its closure.
\end{enumerate}
For the purpose of addressing these issues, we first introduce \textbf{Non-Uniform Growth of the Invariant Set ($\NUGIS$)}. $\NUGIS$, Section~\ref{sec:Nugis}, enables the expansion of the invariant subset with a guarantee so that the limitations of \textbf{Gauranteed Growth} can be overcome. Building on this, our \textbf{Iterative Invariant Set Estimation (IISE)} algorithm, Section~\ref{subseq:IISE}, enhances~\cite{samanipour2024invariant} by using nonlinear $\alpha(x)$ to capture larger sets to tackle \textbf{Linear $\alpha(x)$}, and including boundary points to deal with \textbf{Boundary Point Exclusion}. Our final step is to integrate IISE into \textbf{SEROAISE}, Section~\ref{sec:SEROAISE}, a sequential framework that estimates the RoA efficiently.
\section{Preliminaries}
To begin, we briefly introduce the necessary notations and provide an overview of $\PWA$ functions, barrier functions, and invariant sets.  
\paragraph*{\bf Notation}
Let $S$ be a set. The set of indices corresponding to the elements of $S$ is denoted by $\Mode(S)$. The convex hull, interior, boundary, and closure of $S$ are denoted by $\conv{S}$, $ \Int{S}$, $\partial S$, and $\overline{S}$, respectively. 

For a matrix $A$, $A^T$ denotes its transpose. The infinity norm is denoted by $|\cdot|_\infty$. Additionally, it is important to note that the symbol $\succeq$ has the same element-wise interpretation as $\geq$.

As part of this paper, we present a definition of the $\PWA$ dynamical systems on a partition. As a result, the partition is defined as follows:
\begin{defn}
Throughout this paper, the partition $\mathcal{P}$ is a collection of subsets $\{\pState_i \}_{i \in \Pindex}$, where each $\pState_i$ represents a closed subset of $\mathbb{R}^n$ for all $i\in \Pindex$. In partition $\mathcal{P}$, $\text{Dom}(\mathcal{P})=\cup_{i\in \Pindex}\pState_i$  and $\Int{\pState_i} \cap  \Int{\pState_j} = \emptyset$ for $i\neq j$.
\end{defn}
Another concept that we need is the refinement of a partition.
In mathematical terms, given two partitions $\prtition = \{Y_i\}_{i \in I}$ and $\Rprtition = \{ \qState_j\}_{j \in J}$ of a set $S = \Dom{\prtition} = \Dom{\Rprtition}$, we say that $\Rprtition$ is a refinement of $\prtition$ if $\qState_j \cap Y_i \neq \emptyset$ implies that $\qState_j \subseteq Y_i$.

Furthermore, we use $\mathrm{dim}(X_i)$ to indicate the dimensions of a cell $X_i$ where $i\in I(\prtition)$. 
\subsection{Piecewise Affine Functions}
\label{sec:pwafun}
A piecewise affine function, denoted by $\PWA(x)$, is represented through an explicit parameterization based on a partition $\mathcal{P} = \{\pState_i\}_{i \in \Pindex}$. Each region in the partition corresponds to a set of affine dynamics, described by a collection of matrices $\mathbf{A}_\prtition = \{\genpwaparamM_i \}_{i \in \Pindex}$ and vectors $\mathbf{a}_\prtition = \{\genpwaparamV_i \}_{i \in \Pindex}$. The $\PWA$ function is defined as follows:
\begin{align}\label{eq:definePWAfun}
\PWA(x) &= \genpwaparamM_i \state + \genpwaparamV_i, \quad \text{if } \state \in \pState_i, 
\end{align}
where the region(cell) $\pState_i$ is defined by $n_{h_i}$ hyperplanes as follows:
\begin{align}\label{eq:H rep}
  \pState_i &= \{x \in \mathbb{R}^n \colon \vMatrix_i \state + \vVec_i  \succeq 0\}, 
\end{align}

with matrices $E_i \in \mathbb{R}^{n_{hi} \times n}$ and vectors $e_i \in \mathbb{R}^{n_{hi}}$, which define the boundary hyperplanes of the region $\pState_i$. 
\begin{assum}\label{ass:assumption bounded polytopes}
For the purpose of this research, all cells in partitions are assumed to be bounded polytopes. Therefore, the vertex representation of the $\PWA$ dynamics is applicable. The cell $X_i$ can be represented mathematically as the convex hull of its set of vertices, $\Vertex(X_i)$, as follows:
\begin{equation}
X_i = \text{conv}\{\Vertex(X_i)\}.
\end{equation}
\end{assum}
\subsection{Safety}
To explain the concepts of safety, consider dynamical system~\eqref{eq:general nl}, as locally Lipschitz continuous within the domain $\Domain \subset \mathbb{R}^n$. Based on this assumption, for any initial condition $x_0 \in \Domain$, there is a time interval $I(x_0) = [0, \tau_{\text{max}})$ within which a unique solution $x(t)$ exists, fulfilling the differential equation~\eqref{eq:general nl} and the initial condition $x_0$~\cite{ames2019control}. 
\begin{defn}[Forward invariant set~\cite{ames2019control}]
Let us define a superlevel set $S$ corresponding to a continuously differentiable function $h:\Domain \subset \mathbb{R}^n \rightarrow \mathbb{R}$ for the closed-loop system $f_{cl}$ given by equation~\eqref{eq:general nl} as follows:
\begin{align}\label{eq:BF Domain}
S =& \{x \in \Domain : h(x) \geq 0\},
\end{align}
The set $S$ is considered forward-invariant if the solution $x(t)$ remains in $S$ for all $t \in I(x_0)$ for every $x_0$. If $S$ is forward invariant, the system described by equation~\eqref{eq:general nl} is considered safe with respect to the set $S$.
\end{defn}
\begin{defn}[Barrier function~\cite{ames2019control}]
Let $S \subset \Domain \subseteq \mathbb{R}^n$ represent the superlevel set of a continuously differentiable function $h(x)$. It is said that $h(x)$ is a barrier function if there exists an extended class $\mathcal{K}_\infty$ function $\alpha$ in which:
\begin{equation}\label{eq:barrier general}
L_{f_{cl}} h(x) \geq -\alpha(h(x)), \quad \text{for all } x \in D,
\end{equation}
where lie derivative of $h(x)$ along the closed loop dynamics $f_{cl}$ is denoted by $L_{f_{cl}} h(x)$. 
\end{defn}

\section{Non-Uniform Growth of Invariant Set}
In this section, we introduce the Non-Uniform Growth of Invariant Set ($\NUGIS$) method, which addresses the \textbf{Gauranteed Growth}, challenge~\ref{challenge:Gaurantee}. In order to overcome this limitation, $\NUGIS$ identifies vertices on the boundary of the invariant set that can be incorporated into a larger, newly certified invariant set. $\NUGIS$ has the major advantage of guaranteeing the existence of a larger $\PWA$ invariant set. 
We begin by defining the parameterization of the barrier function, followed by a theorem formalizing $\NUGIS$.
\subsection{Barrier Function Parametrization}
We parameterize the barrier function as a $\PWA$ function on the same partition as the dynamical systems~\eqref{eq:pwa dynamic}.
We expressed the $\PWA$ barrier function and its corresponding invariant set using $h(x,\prtition,\alpha(x))$ and $S(\prtition,\alpha(x))$ to demonstrate their dependence on partition and choosing $\alpha(x)$. It is possible to parameterize the barrier function for the $x \in X_i$ as follows:
\begin{equation}\label{eq: BF PWA}
h_i(x,\prtition,\alpha(x))=s_i^Tx+t_i \quad \text{for} \quad i \in I(\prtition) 
\end{equation}
where $s_i\in \mathbb{R}^n$ and $t_i\in \mathbb{R}$.
It should be noted that $h(x,\prtition,\alpha(x))$ is a continuous function that is differentiable within a cell. 
\begin{assum}\label{assumption:derivate}
Let's consider a cell $X_j$ with local dynamic $\dot{x} = A_jx + a_j$ and a candidate barrier function $h_j(x,\prtition,\alpha(x)) = s_j^T x + t_j$. Then the derivative of the barrier function along the dynamic solution at $\mathbf{x}'\in \Int{X_j}$ can be obtained as follows:
\begin{equation}\label{eq:BF derivative definition}
\dot{h_j}(\mathbf{x}',\prtition,\alpha(x)) = s_j^T(A_j\mathbf{x}'+a_j).
\end{equation}
For more details, please see~\cite{10108069,samanipour2024invariant}.
\end{assum} 
\subsection{NUGIS Theorem}\label{sec:Nugis}
The main idea behind $\NUGIS$ is to expand a $\PWA$ certified invariant set. In light of this, let us assume that $S(\prtition, \alpha(x))$ is a certified invariant set. For a point $\boldsymbol{x}\in \partial{S(\prtition, \alpha(x))}$, we define the following index set:
\begin{equation}\label{eq:index_set_definition}
    I_{\mathcal{P}}(\mathcal{P}, \boldsymbol{x}) = \{i \colon \boldsymbol{x} \in X_i \cap \partial S(\mathcal{P}, \alpha(x))\},
\end{equation}
where $I_{\mathcal{P}}(\mathcal{P}, \boldsymbol{x})$ represents the indices of the cells in the partition $\mathcal{P}$ that include the point $\boldsymbol{x}$.
As noted earlier, the point $\boldsymbol{x}\in \partial S(\prtition,\alpha(x))$ may lie at the intersection of multiple cells in the partition. Consequently, the set $I_{\mathcal{P}}(\mathcal{P}, \boldsymbol{x})$ may contain multiple elements.
\begin{thm}\label{th:LGD}
    Let $\Domain$ be a compact set, and let $S(\prtition,\alpha(x)) = \{ x \in \Domain : h(x,\prtition,\alpha(x)) \geq 0 \}$ represent a $\PWA$ invariant set with respect to the closed-loop dynamics described in~\eqref{eq:pwa dynamic}. 

    If there exist a point $\boldsymbol{x} \in \partial S(\prtition,\alpha(x))$, where the set-valued function $\dot{h}(\boldsymbol{x},\prtition,\alpha(x))$ satisfies the following for an arbitrarily small $\boldsymbol{\epsilon} > 0$
    \begin{equation}\label{eq:NUGIS cond}
        \min_{i \in I(\prtition,\boldsymbol{x})}(\dot{h}_i(\boldsymbol{x},\prtition,\alpha(x))) > \boldsymbol{\epsilon}, i \in I_{P}(\prtition,\boldsymbol{x}), 
    \end{equation}  
    then $\NUGIS$ can be defined as follows:
    \begin{align}
        &\exists X_{new_i},\hspace{0.2 em} \text{s.t}\hspace{0.2 em}  \boldsymbol{x} \in \partial X_{new_i}, X_{new_{i}}\cap \Int{S(\prtition,\alpha(x))}=\emptyset,\nonumber
    \end{align}
    and \begin{equation}
        S(\prtition,\alpha(x))\bigcup_{i\in I_{P}(\prtition,\boldsymbol{x})} X_{new_{i}}=S' \text{ is forward invariant.}
    \end{equation}
\end{thm}
\begin{figure*}
    \centering
    \includegraphics[width=0.8\textwidth]{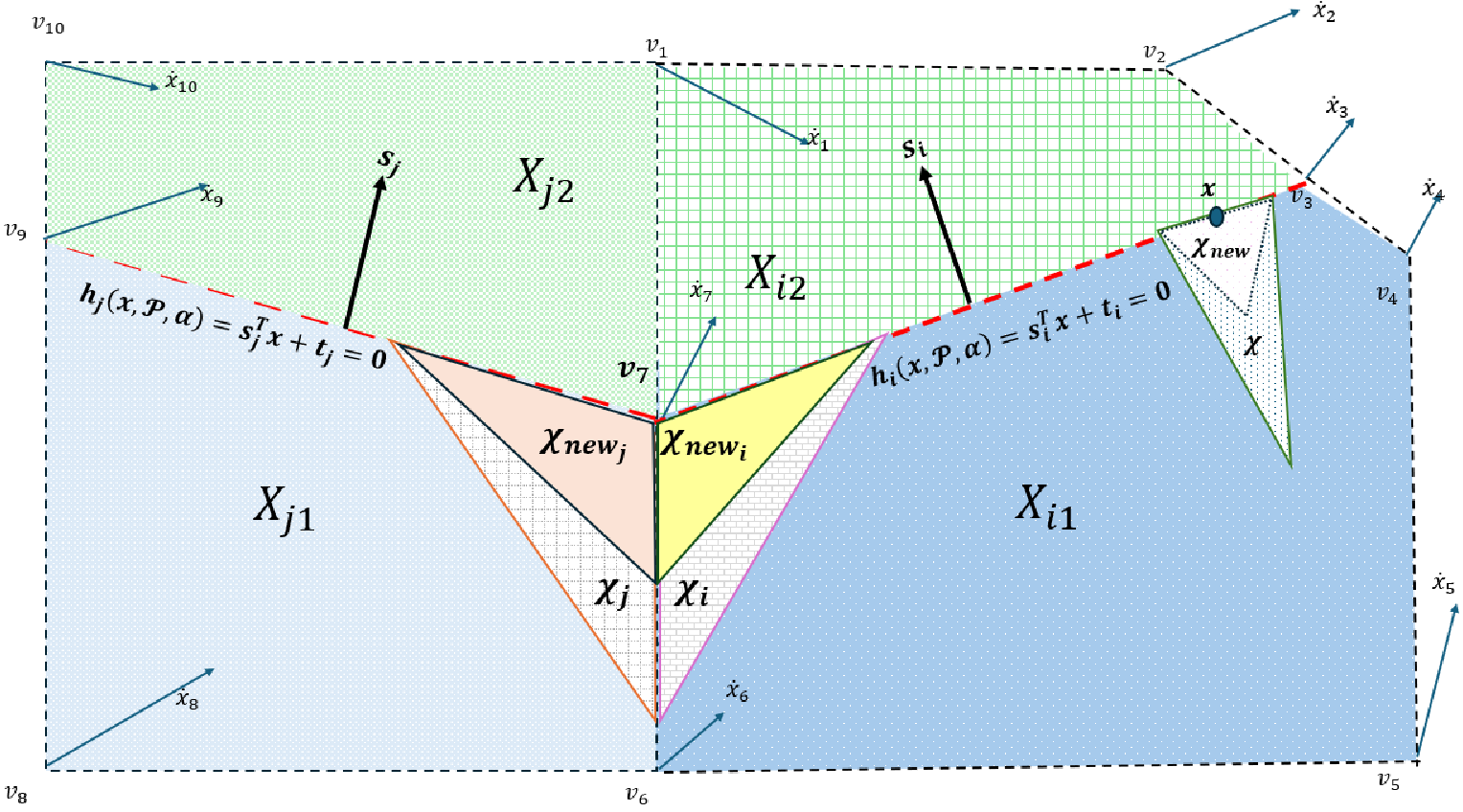}
    \caption{
    Non-uniform Growth of the invariant set for points within the interior of the cells such as $\boldsymbol{x}$ and points on their boundary such as $v_7$ by creating simplex cells as described in Theorem~\ref{th:LGD}. $\dot{x}_i$ represents the vector field at each vertex of these cells. The normal vector of the invariant set toward the origin is $s_i$ for cell $X_i$.}
    \label{fig:LGD}
\end{figure*}
\begin{pf}
The proof proceeds in two parts, addressing: (1) the case where $\dot{h}(x,\prtition,\alpha(x))$ is a singleton function (i.e., \(\boldsymbol{x}\) lies within a single cell), and (2) the case where $\dot{h}(x,\prtition,\alpha(x))$ is set-valued (i.e., \(\boldsymbol{x}\) lies on the boundary of multiple cells). In both cases, we construct new simplex cells around \(\boldsymbol{x}\) and show that incorporating them into the invariant set preserves forward invariance.    
\paragraph*{Step 1: Constructing the Simplex Cell}
Consider the case where $\dot{h}(x,\prtition,\alpha(x))$ is a singleton function at point $\boldsymbol{x}$. Thus, we consider a point $\boldsymbol{x}\in \Int{X_i}$. Cell $X_i$ contains $\partial S(\prtition,\alpha)$ and  $h_i(x,\prtition,\alpha(x))=s_i^Tx+t_i, x\in X_i$ and the following holds. 
\begin{equation}
  \dot{h}_i(\boldsymbol{x},\prtition,\alpha) > \boldsymbol{\epsilon},\quad \boldsymbol{x}\in X_i \nonumber.  
\end{equation}
$\partial S(\prtition,\alpha(x))$ divides the cell $X_i$ into two regions, $X_{i1}$ and $X_{i2}$, where 
\begin{align}\label{eq:dividing cell into 2}
    &X_{i1} = \{ x \in X_i : h(x) \leq 0 \}, \textit{and}\\
    &X_{i2} = \{ x \in X_i : h(x) \geq 0 \}.\nonumber
\end{align}
In Figure~\ref{fig:LGD}, these cells are shown.

Suppose that $(E,e)$ are the hyperplanes of the $X_{i1}$ as described in Equation~\eqref{eq:H rep}. Due to continuity, there exists a simplex polytope $\mathcal{X}$ that is a subset of $X_{i1}$. This polytope is defined by $n+1$ hyperplanes in $\mathbb{R}^n$, and its representation is given by:
\begin{align}\label{eq:X rep}
  \mathcal{X} &= \{x \in X_{i1} \colon \dot{h}_i(x,\prtition,\alpha(x)) \geq \boldsymbol{\epsilon},\hspace{0.5em}  \mathbf{E} \state + \mathbf{e} \succeq 0\},
\end{align}
where $\boldsymbol{x}\in \mathcal{X}$. For clarity, we denote the $k$th hyperplane of $(\mathbf{E}, \mathbf{e})$ by $(\mathbf{E}^k, \mathbf{e}^k)$.

The first hyperplane of $\mathcal{X}$, denoted by $(\mathbf{E}^1, \mathbf{e}^1)$, is aligned with the boundary of invariant set in $X_{i1}$ or in other words $(\mathbf{E}^1,\mathbf{e}^1)=(-s_i^T,-t_i)$ 
as shown in Figure~\ref{fig:LGD}.
Moreover, because of boundedness of the cell $\mathcal{X}$ we can conclude that:
\begin{equation}\label{eq:upper_lower}
    -u^k\leq \mathbf{E}^k(A_ix+a_i)\leq u^k, k=2,3,\dots,n+1
\end{equation}
where $u^k> 0$ is the upper bound of~\eqref{eq:upper_lower}.

Now let us consider another simplex cell $\mathcal{X}_{new_{1}}\subseteq \mathcal{X}$ where $\boldsymbol{x}\in \mathcal{X}_{new_{1}}$. Suppose that its hyperplanes are denoted by $(\mathfrak{E},\mathfrak{e})$ where $\mathfrak{E}\in \mathbb{R}^{(n+1)\times n}$. In convex polytope $\mathcal{X}_{new_{1}}$ we define $\mathfrak{E}^1\in \mathbb{R}^{1\times n}$ and $\mathfrak{E}^1=\mathbf{E}^1$, $\mathfrak{e}^1=\mathbf{e}^1$ and other hyperplanes $(k=2,3,\dots,n+1)$ as follows:
\begin{equation}\label{eq:updated hype}
    \mathfrak{E}^kx+\mathfrak{e}^k=((1-\rho)\mathbf{E}^k+\rho s^T_i)x+\mathfrak{e}^k\succeq0,
\end{equation}
where $\rho \in [0,1)$. Convex polytope $\mathcal{X}_{new_{1}}$ is also shown in Figure~\ref{fig:LGD}. Because $0\leq\rho<1$, the convex polytope $\mathcal{X}_{new_{1}}$ will always have a nonzero volume and it will be the subset of $\mathcal{X}_{new_{1}}$.
The value of $\rho$ for hyperplanes in~\eqref{eq:updated hype} are set as follows:
\begin{equation}\label{eq:rho condition}
    \rho=\max_{k=2,\dots,n+1}\frac{u^k}{u^k+\boldsymbol{\epsilon}}.
\end{equation}
Once the simplex cell has been constructed, it can be checked to see if Forward invariance exists for this set.
\paragraph*{Step 2: Proving Forward Invariance} 
To verify forward invariance of cell $X_{new_{1}}$, we check Nagumo’s condition for hyperplanes $k=2,3,\dots,n+1$:
\begin{align}\label{eq:NG singleton}
    \mathfrak{E}^k\dot{x}=&(1-\rho)\mathbf{E}^k(A_i x + a_i) + \rho s_i^T(A_i x + a_i) \geq\\
    &-(1-\rho)u^k + \rho \boldsymbol{\epsilon} = \nonumber \\
    &\rho(u^k + \boldsymbol{\epsilon}) - u^k\geq \nonumber \\
    &\frac{u^k}{u^k + \boldsymbol{\epsilon}}(u^k + \boldsymbol{\epsilon}) - u^k = 0. \nonumber
\end{align}
The crucial property of $\mathcal{X}_{new_{1}}$ is that any trajectory starting in $\mathcal{X}_{new_{1}}$ will either stay within the cell $\mathcal{X}_{new_{1}}$ or enter the invariant set $S(\prtition,\alpha(x))$. This property follows because the hyperplanes $(\mathfrak{E}^k, \mathfrak{e}^k)$ $(k = 2,3,\dots,n)$ define the boundary of the new invariant set as they satisfy Nagumo's condition.
As a result, we can conclude that $S' = S(\mathcal{P}, \alpha(x)) \cup X_{new_{1}}$ is forward invariant. 

To prove Theorem~\ref{th:LGD} for the set-valued case of $\dot{h}(x,\prtition,\alpha(x))$, we 
follow the same approach to first construct a simplex cell around $\boldsymbol{x}$ and then prove the forward invariance. Before discussing this process in more detail, it is important to note that the following result is valid only in the absence of sliding behavior in the dynamics.

\paragraph*{Step 1:Constructing the Simplex Cells} Consider a point such as $\boldsymbol{x}$, which is located on the boundary of $m$ cells as well as the boundary of the invariant set, for instance, $v_7$, as shown in Figure~\ref{fig:LGD}. In the same way, as in the previous part, the boundary of the invariant set divides each of these $m$ cells into two sub-cells $X_{i1}$ and $X_{i2}$,~\eqref{eq:dividing cell into 2}, for $i=1,2,\dots,m$. The hyperplanes of the cell $X_{i1}, i=1,2,\dots,m$ can be represented by $(E_i,e_i)$. To denote hyperplanes in $X_{i1}$ whose intersection is $\boldsymbol{x}$, we define the following index set.
\begin{equation}
    \mathcal{I}^v_i=\{k\colon E_i^k\boldsymbol{x}+e_i^k=0, E_ix+e_i \succeq 0 \hspace{0.3em}\text{for}\hspace{0.3em} x\in 
    X_{i1}\}.
\end{equation}
By the continuity of the dynamics, in $X_{i1}$ there exist a simplex cell $\mathcal{X}_i \subset X_{i1}$ for $i=1,\dots,m$ such that:  
\begin{align}\label{eq:X_rep}
    \mathcal{X}_i &= \big\{ x \in X_{i1} : \dot{h}_i(x, \prtition, \alpha) \geq \boldsymbol{\epsilon}, \; \mathbf{E}_i \state + \mathbf{e}_i \succeq 0 \big\},
\end{align}  
where $\mathbf{E}_i=[\mathbf{E}_{i1},\mathbf{E}_{i2}]$, $\mathbf{e}_i=[\mathbf{e}_{i1},\mathbf{e}_{i2}]$ and $\mathbf{E}_{i1}=E^{\mathcal{I}^v_i}_i$  and $\mathbf{e}_{i1}=e^{\mathcal{I}^v_i}_i$.
In other words, the first $n$ hyperplanes of $\mathcal{X}_i$ are the same as those in $X_{i1}$ that intersect at the point $\boldsymbol{x}$. 
A simple case in 2D is shown in Figure~\ref{fig:LGD} with two simplices $\mathcal{X}_i$ and $\mathcal{X}_j$ for $v_7$.
 
There exists another simplex cell $X_{new_{i}}\subseteq \mathcal{X}_i$ where $\boldsymbol{x}\in X_{new_{i}}$ for $i=1,\dots,m$. Suppose that $X_{new_{i}}$ hyperplanes are denoted by $(\mathfrak{E}_i,\mathfrak{e}_i)$. We define
$\mathfrak{E}_i=[\mathfrak{E}_{i1},\mathfrak{E}_{i2}]$ and $\mathfrak{e}_i=[\mathfrak{e}_{i1},\mathfrak{e}_{i2}]$ where $\mathfrak{E}_{i1}=\mathbf{E}_{i1}$ and $\mathfrak{e}_1=\mathbf{e}_{i1}$
and $\mathfrak{E}_{i2}$ is the last hyperplane of the simplex cell $X_{new_{i}}$
can be described as follows:
\begin{equation}\label{eq:updated hype_setvalued}
    \mathfrak{E}_{i2}x+\mathfrak{e}_{i2}=((1-\rho)\mathbf{E}_{i2}+\rho s^T_i)x+\mathfrak{e}_{i2}\succeq0.
\end{equation}
where $\rho \in [0,1)$.
\paragraph*{Step 2: Proving Forward Invariance}
To ensure the desired properties, we define $\rho$ as follows:
\begin{equation}\label{eq:rho_condition_set_valued}
    \rho = \max_{i=1,\dots,m} \frac{u_i}{u_i + \boldsymbol{\epsilon}},
\end{equation}
where $-u_i\leq \mathbf{E}_{i2}(A_i x + a_i) \leq u_i$ for $i=1,\dots,m$. Equation~\eqref{eq:rho_condition_set_valued} ensures that Nagumo's condition holds for $(\mathfrak{E}_{i2},\mathfrak{e}_{i2})$ of each $X_{new_{i}}$ same as~\eqref{eq:NG singleton}.
By defining simplex cells $X_{new_{i}}$ for $i = 1, \dots, m$, we construct a new set 
$$S' = S(\mathcal{P}, \alpha(x)) \cup \bigcup_{i=1}^m X_{new_{i}},
$$
where the boundaries of $S'$ are either $\partial S(\mathcal{P}, \alpha(x))$ or $\mathfrak{E}_{i2}$ for $i = 1, \dots, m$. Consequently, the set $S'$ is forward invariant, as Nagumo's condition holds for both $\partial S(\mathcal{P}, \alpha(x))$ and $\mathfrak{E}_{i2}$ for all $i = 1, \dots, m$.

As a result, we proved that there is a forward invariant set that contains point $\boldsymbol{x}$ in its interior if~\eqref{eq:NUGIS cond} holds.
\end{pf}
This theorem establishes that if the condition~\eqref{eq:NUGIS cond} is satisfied for a point $\boldsymbol{x}$ on the boundary of the $\PWA$ invariant set, then there exists a larger invariant set that includes $\boldsymbol{x}$ in its interior. The $\NUGIS$ Theorem guarantees the existence of a larger  $\PWA$ invariant set and constructs this larger invariant set as described in the Theorem~\ref{th:LGD}. However, this new larger invariant set will not be directly utilized. Instead, the following section will explain how this theorem contributes to subsequent developments. 
\section{Iterative Invariant Set Estimator(IISE)}\label{subseq:IISE}
We introduce the \emph{Iterative Invariant Set Estimator} (IISE) to construct larger, less conservative invariant sets for the $\PWA$ system in~\eqref{eq:pwa dynamic}, using the $\NUGIS$ framework from Theorem~\ref{th:LGD}. For a certified $\PWA$ invariant set, IISE retains all interior vertices within that set while expanding it non-uniformly through $\NUGIS$. The search for a strictly larger invariant set is possible given the results of $\NUGIS$ and overcome the challenge of \textbf{Guaranteed Growth}~\ref{challenge:Gaurantee}. To provide more flexibility than using \textbf{Linear $\alpha$} in equation~\eqref{eq:barrier general}, IISE applies a Leaky ReLU function to the barrier function constraint~\eqref{eq:barrier general}. Furthermore, IISE addresses the \textbf{Boundary Points Exclusion}, challenge~\ref{goal: BPI},  by explicitly incorporating points in $\overline{\Domain}$. 

The IISE approach categorizes all vertices in the domain into distinct groups with respect to the certified $\PWA$ invariant set, which guides the iterative expansion process. The following sections outline these categories, formulate an optimization problem based on them, and present the complete IISE algorithm.
\subsection{Categorizing Vertices}\label{subsec:cat vert}
To enable the Iterative Invariant Set Estimator (IISE) to retain interior vertices and expand the invariant set non-uniformly via $\NUGIS$ (Theorem~\ref{th:LGD}), we categorize all vertices of a certified $\PWA$ invariant set into five distinct groups. By categorizing vertices, we can determine which vertices to preserve, grow toward, or exclude during the optimization process. Prior to defining these categories, we establish preliminary concepts.

IISE operates iteratively, with \(m\) denoting the current iteration. At iteration $m+1$, we assume a certified invariant set $S(\prtition_m, \alpha_m(x))$ from iteration $m$ for the dynamic~\eqref{eq:pwa dynamic}, with a corresponding barrier function \(h(x, \prtition_m, \alpha_m(x))\). To align the invariant set’s boundary with cell boundaries, we refine the partition \(\prtition_m\) into:
\begin{equation}\label{eq:refined partition}
\prtition_m^* = \{ X_k \}_{k \in I(\prtition_m^*)},
\end{equation}
where each cell \(X_k \in \prtition_m^*\) is:
\begin{align}
X_k =
\begin{cases}
X_{i_1} \text{ and } X_{i_2}, & \text{if } X_i \cap \partial S \neq \emptyset, \\
X_i, & \text{otherwise},
\end{cases}
\end{align}
where $i \in I(\prtition_m)$ and subregions:
\begin{align*}
X_{i_1} &= X_i \cap S(\prtition_m, \alpha_m(x)), \\
X_{i_2} &= X_i \setminus \Int{S(\prtition_m, \alpha_m(x))}.
\end{align*}
This ensures \(\partial S(\prtition_m, \alpha_m(x))\) coincides with cell boundaries in \(\prtition_m^*\).

We define the set of all vertex-cell pairs in the refined partition:
\begin{equation}\label{eq:all pairs}
I_{\Domain}(\prtition_m^*) = \{ (i,k) : i \in I(\prtition_m^*), v_k \in \Vertex(X_i) \},
\end{equation}
and identify boundary vertices of \(\Domain\):
\begin{equation}
I_b(\prtition_m^*) = \{ (i,k) \in I_{\Domain}(\prtition_m^*) : v_k \in \partial \Domain, X_i \cap \partial \Domain \neq \emptyset \}.
\end{equation}
Unlike~\cite{samanipour2024invariant}, which excludes all $v_k$ where $(i,k) \in I_b(\prtition_m^*)$ from the invariant set, $S \subseteq \Int{\Domain}$, IISE allows $S \subseteq \overline{\Domain}$ via the following proposition.

\begin{prop}\label{prop:BPI}
For a boundary vertex \(v_j \in X_i\), \((i,j) \in I_b(\prtition_m^*)\), with dynamics \(\dot{x}_i = A_i x + a_i\) and a hyperplane \((E', e')\) such that \(E'x + e' = 0\) for \(x \in X_i \cap \partial \Domain\), if \(E'(A_i v_j + a_i) < 0\), then \(v_j\) cannot belong to any invariant set \(S \subseteq \overline{\Domain}\).
\end{prop}
\begin{pf}
We can prove this proposition from a geometric perspective. The hyperplane defined by $(E', e')$, as described in equation~\eqref{eq:H rep}, has $E'$ as its normal vector pointing towards the origin. If the inner product of the vector field at $v_j$ and $E'$ is negative, the trajectory will cross the hyperplane and exit the domain $\Domain$. Thus, no invariant set can contain $v_j$. It is important to emphasize that this proposition applies only to boundary vertices.
\end{pf}
\begin{figure}
    \centering
    \includegraphics[width=0.85\linewidth]{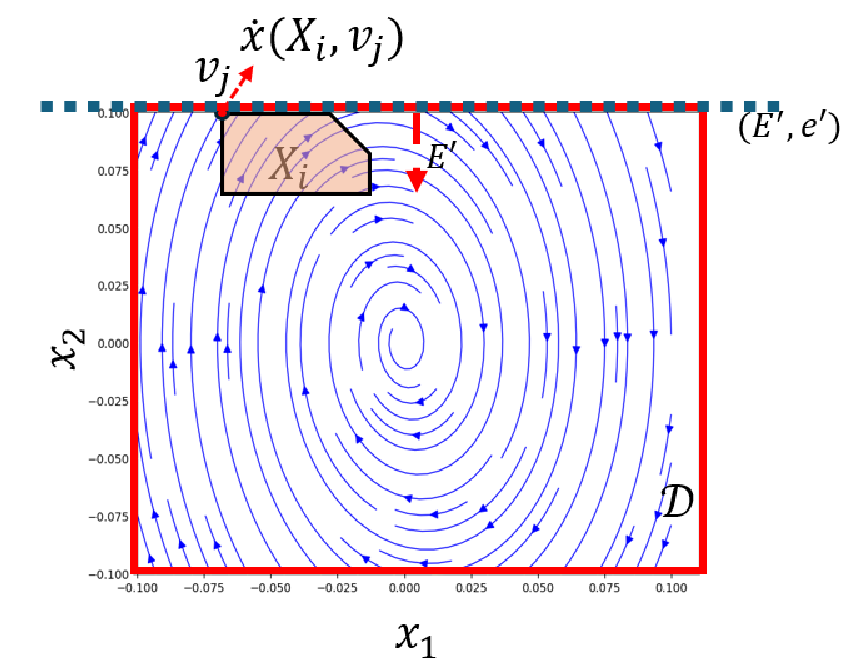}
    \caption{If \(E' \dot{x}(v_j) < 0\), the vertex \(v_j\) cannot be included in any invariant set \(S \subseteq \overline{\Domain}\).}
    \label{fig:prop2}
\end{figure}

\begin{figure}
    \centering
    \includegraphics[width=0.75\linewidth]{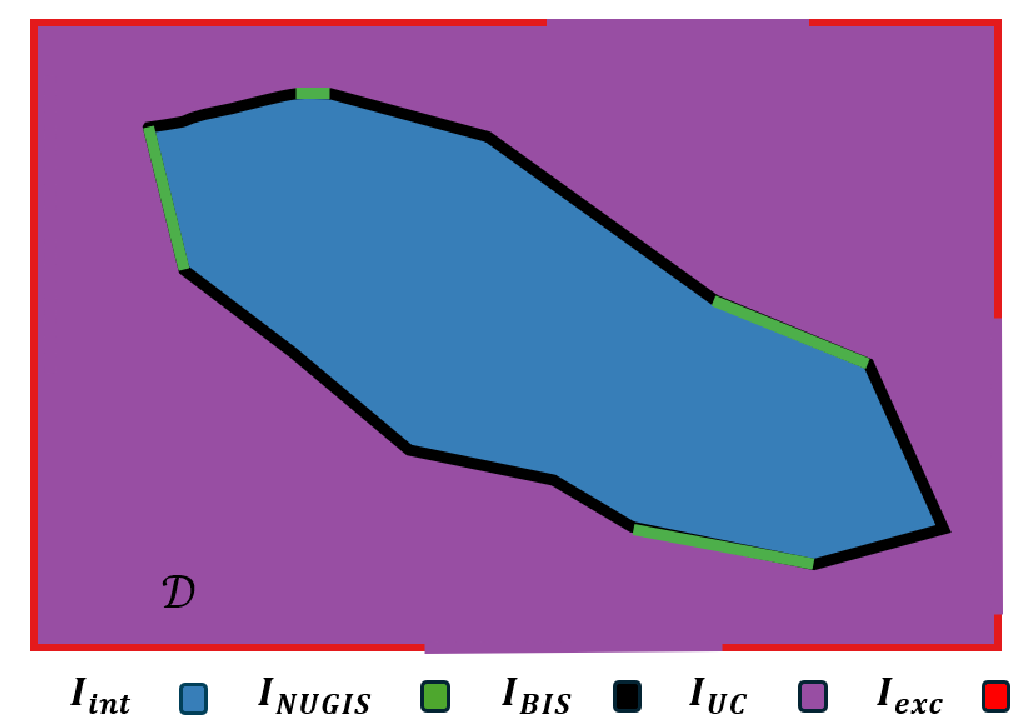}
    \caption{Vertex categorization for IISE, where blue region denotes a $\PWA$ certified invariant set interior, green segments show $\NUGIS$ points, black segments show points that could stay on the boundary of the invariant set, and the red border marks the exclusion zone. The purple denotes the points that refer to the $I_{UC}$ as described in category~\ref{cat5:Unclass}. }
    \label{fig:vert_cat}
\end{figure}

With these preliminaries, we categorize the vertices as follows:
\begin{enumerate}
    \item \label{cat1:int}The first Category consists of pairs of cells and their corresponding vertices that lie strictly within the interior of the invariant set $S(\prtition_m,\alpha_m(x))$. These points are mathematically defined as:
    \begin{align}\label{eq:I_int}
        I_{\mathrm{int}}(\prtition_m^*)& = \{(i, k) : h_i(v_k, \prtition_m^*, \alpha_m(x)) > 0, \\&\; i \in I(\prtition_m^*) \}.\nonumber
    \end{align}
 Throughout the algorithm, these points are guaranteed to remain within the interior of the invariant set.
    \item \label{cat2:LGD}The second category, denoted by $I_{\text{\NUGIS}}(\prtition_m^*)$, includes points on the boundary of the invariant set $S(\prtition_m,\alpha_m(x))$. The algorithm aims to include these points in the interior of the new invariant set, ensuring they are no longer boundary points. Set $I_{\text{\NUGIS}}(\prtition_m^*)$ can be expressed mathematically as follows:
\begin{align}
  I_{\NUGIS}(\prtition_m^*)&=
  \{(i,k) \in I_{\partial S}(\prtition_m^*):\\ &\min_{i}(\dot h_i(v_k,\prtition_m^*,\alpha_m(x)))>\boldsymbol{\epsilon}\nonumber\},
\end{align}
where:    
\begin{align}
 I_{\partial S}(\prtition_m^*)=&\{(i,k)\colon
 h_i(v_k,\prtition_m^*,\alpha_m(x))=0\nonumber\}.
\end{align}
    \item \label{cat3:NLGD} The third category, $I_{\mathrm{BIS}}(\prtition_m^*)$, consists of points that could remain on the boundary of the invariant set as follows:
    \begin{align}\label{eq: BIS points}
     I_{\mathrm{BIS}}(\prtition_m^*)= I_{\partial S}(\prtition_m^*)\setminus I_{\NUGIS}(\prtition_m^*),
    \end{align}
        \item \label{cat4:Ninv}The fourth category, denoted by $I_{\text{excl}}(\prtition_m^*)$, includes vertices that must remain outside the invariant set. These points are specified as non-invariant as a result of Proposition~\ref{prop:BPI} and can be expressed as follows:
    \begin{align}\label{eq:non invariant points}
&I_{\text{excl}}(\mathcal{P}_m^*) = \{ (i, k) \in I_b(\mathcal{P}_m^*) : E'x + e' = 0,\\&  x \in X_i \cap \partial \Domain, \quad E' (A_i v_k + a_i) < 0 \nonumber\}.
\end{align}
    \item\label{cat5:Unclass}The final category, $I_{\text{UC}}(\prtition_m^*)$, includes vertices that do not belong to any of the above categories. Although we aim to keep them within the interior of the invariant set, some may not satisfy the conditions to remain inside. We formally define this set as
    \begin{align}\label{eq:UC}
    &I_{UC}(\prtition_m^*) = I_{\Domain}(\prtition_m^*) \setminus (I_{\text{int}}(\prtition_m^*) \cup I_{\text{excl}}(\prtition_m^*)\nonumber\\
    &\cup I_{\text{\NUGIS}}(\prtition_m^*) \cup I_{\mathrm{BIS}}(\prtition_m^*) ),
    \end{align}
    where $I_{\Domain}(\prtition_m^*)$ defined in~\eqref{eq:all pairs} .
\end{enumerate}
These categories of points are shown in Figure~\ref{fig:vert_cat}. As a result of these categories, IISE is able to grow systematically in subsequent optimization steps. In the following section, we will discuss the details.
\subsection{Forming Linear Optimization Problem}
Building on the categorization of vertices in Section~\ref{subsec:cat vert}, we now formulate a linear optimization problem to determine the $\PWA$ invariant set, ensuring compatibility with the goals of each vertex category. However, selecting an appropriate \(\alpha(x)\) remains a challenge to the proposed method.

\begin{figure}
    \centering
    \includegraphics[width=0.8\linewidth]{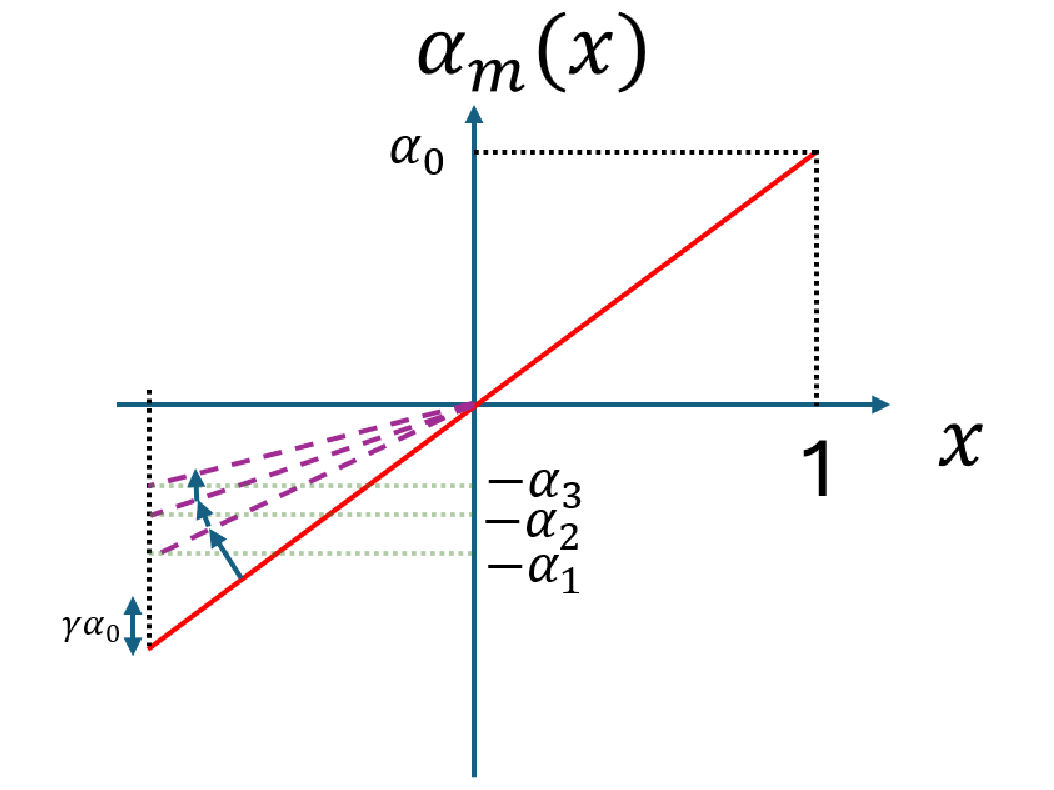}
    \caption{The process of updating the $\mathcal{K}_{\infty}$ function $\alpha_m(x)$ in each iteration with learning rate $\gamma$.}
    \label{fig:updating alpha}
\end{figure}

In order to overcome \textbf{Linear $\alpha(x)$} challenge as described in section~\ref{goal:linear alpha}, we utilize an updated $\mathcal{K}_{\infty}$ function as follows: 
\begin{equation}\label{eq:updated Leaky}
    \alpha_m(x)=\alpha_0\sigma_{\left(\frac{\alpha_m}{\alpha_0}\right)}(x), \quad m=1,2,3,\dots
\end{equation}
where $\alpha_0>0$ and $\sigma_{\left(\frac{\alpha_m}{\alpha_0}\right)}$ is a Leaky ReLU function defined as 
    \begin{equation}\label{eq:Leaky relu}
    \sigma_{\left(\frac{\alpha_m}{\alpha_0}\right)}(x) =
    \begin{cases} 
    x, & \text{if } x \geq 0, \\
    \left(\frac{\alpha_m}{\alpha_0}\right)x, & \text{if } x < 0,
    \end{cases}
    \end{equation}
and $\alpha_m=(1-\gamma )\alpha_{m-1}$ and $\gamma<1$ is the learning rate. The process of updating the $\alpha_m(x)$ in each iteration is depicted in Figure~\ref{fig:updating alpha}.
The feasibility of using Leaky ReLU $\mathcal{K}_\infty$ function is discussed in~\cite{samanipour2025replacing}[Theorem 1].

Now we can construct the new optimization problem using five categories of points and leaky ReLU function~\eqref{eq:updated Leaky} upon a known barrier function $h(x,\prtition^*_{m},\alpha_{m}(x))$. The cost function, which penalizes deviations via slack variables, is defined as:
\begin{equation}\label{eq:cost seq}
\mathcal{J}=\Gamma(\sum_{i=1}^{M}\tau_{b_{i}}+\tau_{\mathrm{int}}+\tau_{\NUGIS}+\tau_{\mathrm{BIS}})+\sum_{i=1}^{N}\tau_{UC_{i}}.
\end{equation}
where $M$ and $N$ denote the number of cells in $I_{\mathrm{excl}}(\prtition^*_m)$ and $I_{UC}(\prtition^*_m)$, respectively. Moreover, $\Gamma>1$ is the penalty for non-zero slack variables. $\tau_b,\tau_{\mathrm{int}},\tau_{\NUGIS}$ and $\tau_{\mathrm{BIS}}$ are the slack variables for category~\ref{cat4:Ninv}, category~\ref{cat1:int}, category~\ref{cat2:LGD} and category~\ref{cat3:NLGD} respectively. $\tau_{UC}$ is the slack variable for the points in category~\ref{cat5:Unclass}.
The final optimization problem is constructed at iteration 
$m$ as follows :
\allowdisplaybreaks
\begin{subequations}\label{eq:opt seq}
\begin{align} 
&\min_{ s_i, t_i,\tau_{\mathrm{int}},\tau_{b_{i}},\tau_{UC_{i}},\tau_{\NUGIS},\tau_{\mathrm{BIS}}}\mathcal{J} ,\nonumber\\
&\text{Subject to:} \nonumber\\
&h_i(v_k,\prtition^*_m,\alpha_m(x))-\tau_{b_i}\leq -\epsilon_1,\forall (i,k) \in I_{\text{excl}}(\prtition^*_{m-1})\label{eq:NI seq}\\
&h_i(v_k,\prtition^*_m,\alpha_m(x))+\tau_{\mathrm{BIS}}\geq 0, \hspace{0.2em}\forall (i,k) \in I_{\mathrm{BIS}}(\prtition^*_{m-1})\label{eq:NLGD seq}\\
&h_i(v_k,\prtition^*_m,\alpha_m(x)))+\tau_{\mathrm{int}}\geq \epsilon_3,\forall (i,k) \in I_{\mathrm{int}}(\prtition^*_{m-1})\label{eq:interior seq}\\
&h_i(v_k,\prtition^*_m,\alpha_m(x))+\tau_{\NUGIS}\geq \epsilon_3,\nonumber\\ &\forall (i,k) \in I_{\NUGIS}(\prtition^*_{m-1})\label{eq:LGD seq}\\
&h_i(v_k,\prtition^*_m,\alpha_m(x))+\tau_{UC_{i}}\geq \epsilon_2, \hspace{0.2em}\forall (i,k) \in I_{UC}(\prtition^*_{m-1})\label{eq:UC seq}\\
&\dot{h}(v_k,\prtition^*_m,\alpha_m(x))+\alpha_0 h(v_k,\prtition^*_m,\alpha_m(x))\geq \epsilon_3,\label{eq:nag seq int}\\
&\forall (i,k) \in I_{\mathrm{int}}(\prtition^*_m)\cup I_{\NUGIS}(\prtition^*_m)\cup I_{\mathrm{BIS}}(\prtition^*_m)\nonumber\\
&\dot{h}(v_k,\prtition^*_m,\alpha_m(x))+\alpha_m h(v_k,\prtition^*_m,\alpha_m(x))\geq \epsilon_3,\label{eq:nag seq out}\\
&\forall (i,k) \in I_{\mathrm{excl}}(\prtition^*_m)\cup I_{UC}(\prtition^*_m)\nonumber\\
&h_i(v_k,\prtition^*_m,\alpha_m(x))=h_j(v_k,\prtition^*_m,\alpha_m(x)), \\ &\forall v_k\in \Vertex(X_i)\cap \Vertex(X_j)\nonumber\\
&\tau_{b_i},\tau_{int},\tau_{\NUGIS},\tau_{\mathrm{BIS}},\tau_{UC_{i}}\geq 0,\label{eq:PS seq}
\end{align}
\end{subequations}
As part of this optimization problem, we impose constraints to ensure that all vertices in $I_{\mathrm{excl}}(\prtition^*_m)$ lie outside the invariant set, as specified by constraint~\eqref{eq:NI seq}. Furthermore, constraint~\eqref{eq:NLGD seq} guarantees that points in $I_{\mathrm{BIS}}(\prtition^*_{m-1})$ lie either on the boundary or within the interior of $S(\prtition^*_m, \alpha_m(x))$.
We further aim to enforce the invariant set to include all points from the invariant set $S(\prtition^*_{m-1}, \alpha_{m-1}(x))$, as defined by constraint~\eqref{eq:interior seq}. Moreover, the new invariant set can incorporate points that lie on the boundary of the previous invariant set $S(\prtition^*_{m-1}, \alpha_{m-1}(x))$ into the interior of $S(\prtition^*_m, \alpha_m(x))$, as governed by constraint~\eqref{eq:LGD seq} for points in $I_{\text{\NUGIS}}(\prtition^*_{m-1})$.
Finally, constraints~\eqref{eq:nag seq int} and~\eqref{eq:nag seq out} ensure that the condition specified in~\eqref{eq:barrier general} is satisfied for all points in $\Domain$ under a Leaky ReLU parameterization $\alpha_m(x)$. Solving the optimization problem with a Leaky ReLU parameterization typically requires mixed-integer programming, categorizing points into five distinct groups allows the use of linear optimization instead. Specifically, when a point lies within the invariant set or is intended to be included in the invariant set, we use $\alpha_0$. For points belonging to the categories $I_{\text{excl}}(\prtition^*_m)$ and $I_{UC}(\prtition^*_m)$, we use $\alpha_m$ instead.
\begin{lem}
Optimization problem~\eqref{eq:opt seq} is always feasible.
\end{lem}
\begin{pf}
Due to the definition of slack variables, the optimization problem is always feasible. 
For more information, please see~\cite{samanipour2024invariant}.
\end{pf}

\begin{rem}
Although optimization problem~\eqref{eq:opt seq} is always feasible, the solution is certified invariant set if $$\sum_{i=1}^{M}\tau_{b_{i}}+\tau_{\mathrm{int}}+\tau_{\NUGIS}+\tau_{\mathrm{BIS}}=0.$$
There may be a need for refinement if a certified invariant set cannot be obtained. The vector field refinement technique~\cite{10313502} needs to be applied to all cells with $\tau_{b_i}\neq 0$, $\tau_{\NUGIS}\neq 0$, or $\tau_{\mathrm{BIS}}\neq 0$.
\end{rem}
By applying the new optimization problem~\eqref{eq:opt seq}, the invariant set non-uniformly expands to include vertices within $I_{\NUGIS}(\prtition_m^*)$ at each iteration.  
\begin{lem}\label{lemma:construct two inv}
    Consider the invariant set $S(\prtition_{m-1}, \alpha_{m-1}(x))$ as a certified $\PWA$ invariant set with respect to~\eqref{eq:pwa dynamic}. If a certified invariant set is obtained from the optimization problem~\eqref{eq:opt seq}, $S(\prtition_{m},\alpha_m(x))$, where $\mathcal{P}_m =\mathrm{Ref}(\prtition^*_{m-1})$, then $S(\mathcal{P}_{m-1}, \alpha_{m-1}(x)) \subset S(\mathcal{P}_m, \alpha_m(x))$.
\end{lem}
\begin{pf}
    By construction of the optimization problem~\eqref{eq:opt seq}, it follows that
    $$
    \sum_{i=1}^{M}\tau_{b_{i}} + \tau_{\mathrm{int}} + \tau_{\NUGIS} + \tau_{\mathrm{BIS}} = 0,
    $$
    if the solution is a certified invariant set. Therefore, all points within $S(\prtition_{m-1}, \alpha_{m-1}(x))$ are contained in the refined invariant set $S(\prtition_{m}, \alpha_{m}(x))$ according to~\eqref{eq:interior seq} and~\eqref{eq:NLGD seq}. Moreover, because $\tau_{\NUGIS} = 0$, additional points will be included in $S(\prtition_m, \alpha_m(x))$. Therefore, $S(\mathcal{P}_{m-1}, \alpha_{m-1}(x)) \subset S(\mathcal{P}_m,\alpha_m(x))$.
\end{pf}
\subsection{Algorithm Description and Termination Condition}\label{subsec:IISE termination}
Following the formulation of the linear optimization problem in the previous section, we now introduce the Iterative Invariant Set Estimator (IISE) algorithm to iteratively compute the $\PWA$ invariant set. The details of this algorithm are presented in Algorithm~\ref{alg:Sequential Invariant Set Estimator(IISE)}. 
The process begins with a barrier function obtained from~\cite{samanipour2024invariant}. The algorithm may begin with any $\PWA$ invariant set rather than the invariant set obtained by using~\cite{samanipour2024invariant}. Then. the algorithm updates the $\mathcal{K}_{\infty}$ function $\alpha_m(x)$ using the Leaky ReLU parameterization from~\eqref{eq:updated Leaky}.  The inner loop of the algorithm is solving the optimization problem~\eqref{eq:opt seq} according to the vertices categories. Whenever the optimization problem fails to find a certified invariant set, vector-field refinement~\cite{10313502} is employed. 
The outer loop is repeated until the $\NUGIS$ condition,~\eqref{eq:NUGIS cond}, does not hold or the maximum number of iterations has been reached. 

The termination condition is inspired by Zubov’s theorem, where the Zubov function $W(x)$ satisfies $W(x) = 1$ at the boundary, and $\lim_{W(x) \rightarrow 1} \dot{W}(x) = 0$ characterizes the Region of Attraction (RoA). In our method, we update \(\alpha_m(x)\) as shown in~\eqref{eq:updated Leaky} to relax the barrier function condition~\eqref{eq:barrier general} for points outside the invariant set. For \(x \notin S(\prtition^*_m, \alpha_m(x))\), the barrier function condition requires that \(\dot{h}(x, \prtition^*_m, \alpha_m(x))\) is strictly positive. By employing a smaller \(\alpha_m(x)\), however, we allow these positive values to be only marginally above zero. This means that, although \(\dot{h}(x, \prtition^*_m, \alpha_m(x))\) must remain strictly positive outside the invariant set, the use of a smaller \(\alpha_m(x)\) ensures that the values can be made small, thus closely approximating the boundary behavior predicted by Zubov's theorem.
$\boldsymbol{\epsilon}$ controlling the proximity to the maximal invariant set. A smaller \(\boldsymbol{\epsilon}\) yields a tighter approximation but increases computational complexity.

  
The last note that we need to add is that as a result of Lemma~\ref{lemma:construct two inv}, the resulting invariant sets satisfy $S(\prtition_1, \alpha_1(x)) \subset S(\prtition_2, \alpha_2(x)) \subset \dots \subset S(\prtition_m, \alpha_m(x))$. It is evident that as $m$ increases, the obtained invariant set becomes larger since the invariant set grows monotonically. However, it is important to note that increasing $m$ may lead to higher computational costs.

In this algorithm, we addressed the challenge of \textbf{Guaranteed Growth} by using the concept of $\NUGIS$. Moreover, we showed that IISE is capable of including points on the boundary of the pre-selected $\Domain$, as a result of Proposition~\ref{prop:BPI}. Also, we utilized the Leaky ReLU in the barrier function constraint to address the challenge of \textbf{Linear $\alpha(x)$}. Therefore, all challenges have been overcome. The following example illustrates the performance of the algorithm.

\begin{algorithm}[tb]
    \begin{algorithmic}
    \REQUIRE  $\PWA(x)$ dynamic~\eqref{eq:pwa dynamic}, a barrier function $h(x,\prtition_0,\alpha_0(x))$ where $\alpha_0(x)=\alpha_0 x$ obtained using~\cite{samanipour2024invariant}.
    \STATE Initialize \(m = 0\) and \(\gamma < 1\).
    \STATE Set all the slack variables in~\eqref{eq:cost seq} $\tau_b=\tau_{int}=\tau_{\NUGIS}=\tau_{\overline{\NUGIS}}=\tau_{UC}\rightarrow \infty$.
    \WHILE{$\min_{i \in I(\prtition_m,\boldsymbol{x})}(\dot{h}_i(\boldsymbol{x},\prtition_m,\alpha_m(x))) > \boldsymbol{\epsilon}$ and $m\leq \textit{max-iter}$}
        \STATE Refine the partition $\prtition_m$  to obtain $\prtition_m^*$ as described in~\eqref{eq:refined partition}.
        \STATE Find $I_{\mathrm{int}}(\prtition^*_m)$, $I_{UC}(\prtition^*_m)$, $I_{\NUGIS}(\prtition^*_m)$, $I_{\mathrm{BIS}}(\prtition^*_m)$ and $I_b(\prtition^*_m)$.
        \STATE $m=m+1$.
        \STATE Update $\alpha_m(x)$ as described in~\eqref{eq:updated Leaky}.
        \WHILE{$\sum_{i=1}^{M}\tau_{b_{i}}+\tau_{\mathrm{int}}+\tau_{\NUGIS}+\tau_{\mathrm{BIS}}\neq 0$}
        \STATE{Search for a certified invariant set using optimization problem~\eqref{eq:opt seq}  for the $\PWA$ dynamics~\eqref{eq:pwa dynamic} with $\alpha_m(x)$ over the refined partition of the domain.}
            \FOR{each $i$ such that $\tau_{b_i} \neq 0$, $\tau_{\NUGIS}\neq 0$ and$\tau_{\mathrm{BIS}}\neq 0$ and $\tau_{\mathrm{int}}\neq 0$} 
                \STATE{Refine corresponding cells~\cite{10313502}.}
            \ENDFOR 
        \ENDWHILE
    \ENDWHILE
    \RETURN $S(\prtition_m,\alpha_m)$.  
    \end{algorithmic}
    \caption{Iterative Invariant Set Estimator(IISE) for $\PWA$ dynamics~\eqref{eq:pwa dynamic} as described in section~\ref{subseq:IISE}}
    \label{alg:Sequential Invariant Set Estimator(IISE)}    
\end{algorithm}
\begin{exmp}[Inverted Pendulum~\cite{rabiee2023soft}]\label{example:IP}
The inverted pendulum system can be modeled by the following equations~\cite{rabiee2023soft}:
\begin{align}
\dot{x}_1 &= x_2 \nonumber\\
\dot{x}_2 &= \sin(x_1) + u \nonumber
\end{align}
where $x_1$ represents the pendulum's angle, and $x_2$ is its angular velocity. The control input $u$, defined as $u = -3x_1 - 3x_2$, is constrained by a saturation limit between the lower bound of $-1.5$ and the upper bound of $1.5$.
The system's domain, $\Domain$, is given by $\Domain = \pi - |x|_\infty \geq 0$. We utilize a ReLU neural network with a single hidden layer with eight neurons to approximate the right-hand side of the dynamics. First, we used the proposed method in~\cite{samanipour2024invariant} to provide a certified $\PWA$ invariant set. Then, we estimate the invariant set using IISE in 3 iterations. The results are shown in Figure~\ref{fig:IISE}. Comparably, the certified invariant set is significantly larger than the known invariant set obtained from the algorithm proposed in~\cite{samanipour2024invariant}. The invariant set grows monotonically with each iteration, as shown in Figure~\ref{fig:IISE}. 
\begin{figure}
    \centering
    \includegraphics[width=1.1\linewidth]{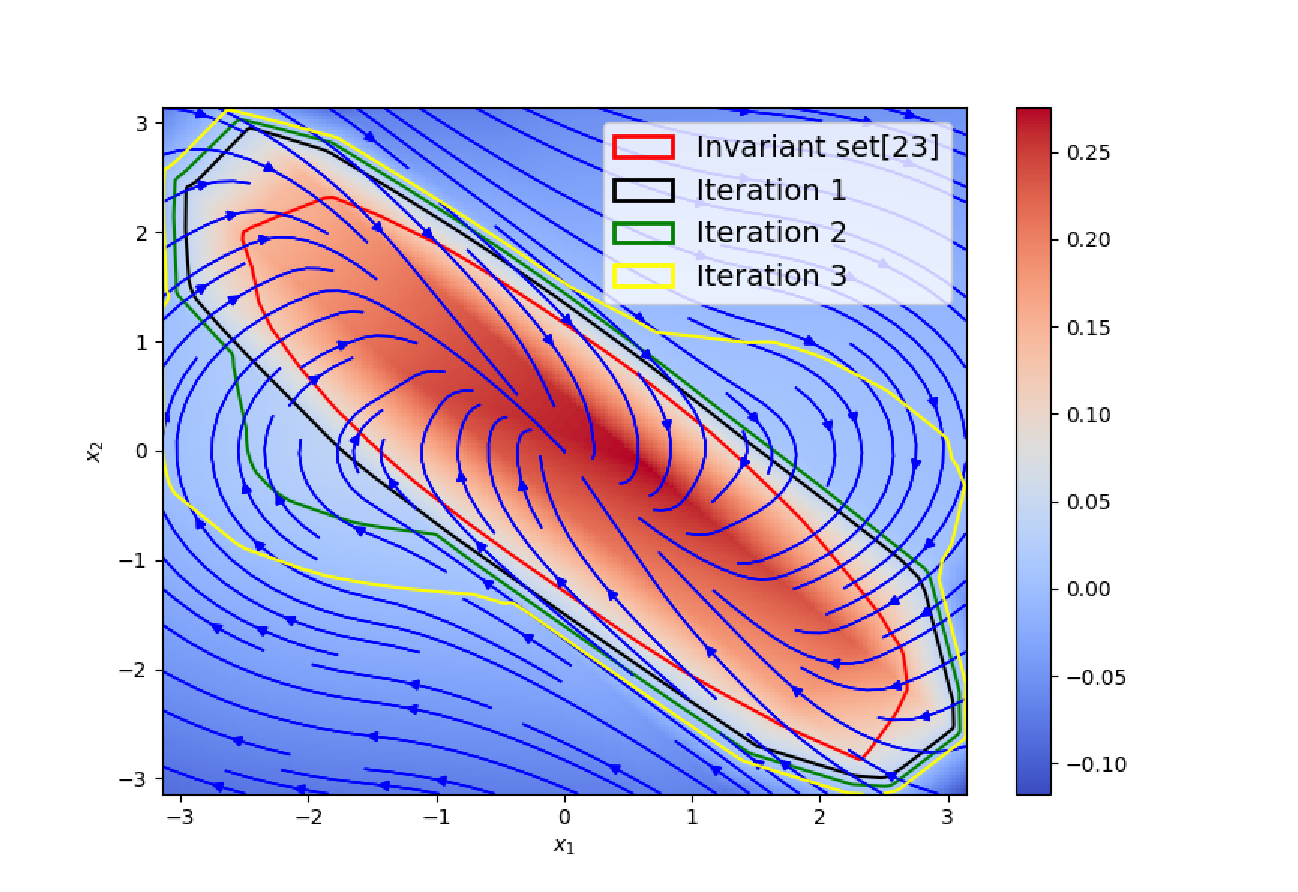}
    \caption {
    Using IISE to expand the certified invariant set obtained using~\cite{samanipour2024invariant} in 3 iterations for the ReLU NN used to identify the dynamic presented in Example~\ref{example:IP}. As can be seen in each step, the estimated invariant set will increase monotonically.}
    \label{fig:IISE}
\end{figure}
\end{exmp}
\section{SEROAISE}\label{sec:SEROAISE}
After addressing the Challenges associated with obtaining the certified invariant set using the IISE algorithm, we can proceed to present the RoA Estimation Algorithm.
This section presents SEROAISE as a sequential approach to achieving the RoA for the $\PWA$ dynamics. First, we determine a certified invariant set within a predefined domain using the IISE method, an advancement over the technique presented in~\cite{samanipour2024invariant} that yields a larger invariant set. 
There is a detailed description of the IISE algorithm in Algorithm~\ref{alg:Sequential Invariant Set Estimator(IISE)}.
Once the invariant set is obtained, the second step involves constructing a $\PWA$ Lyapunov-like function to ensure that the estimated invariant subset is the RoA. This process is divided into the following stages:
\paragraph*{Partition Redefinition}
The partition is redefined to include cells within the invariant set. This is formalized as:
\begin{equation}\label{eq:New partition}
I(\prtition') = \{ i : h_i(x) \geq 0, x \in X_i, i \in I(\prtition^*) \}.
\end{equation}

\paragraph*{Function Definition}
A candidate $\PWA$ Lyapunov-like function $V(x)$ is defined over $\prtition'$ as:
\begin{align}\label{Eq:Lyap PWA}
    V(x) = 
    \begin{cases} 
        p_i^T x + q_i & \text{for } i \in I_1(\prtition'), \\
        p_i^T x & \text{for } i \in I_0(\prtition'),
    \end{cases}
\end{align}
where \(I_1(\prtition')\) includes cells not containing the origin, and \(I_0(\prtition')\) includes cells containing the origin.

\paragraph*{Optimization Problem}
To find \(V(x)\), we formulate the following optimization problem:
\begin{subequations}\label{eq:Lyap-lik opt}
\begin{align}
    \min_{p_i, q_i, \tau_i} \quad & \sum_{i=1}^{N} \tau_i, \label{eq:cost lyap like} \\
    \text{Subject to:} & \nonumber \\
    \dot{V}_i(v_k) - \tau_i \leq -\epsilon_1, & \forall i \in I(\prtition'), v_k \in \Vertex(X_i) \setminus \{0\}, \label{eq:neg der} \\
    V_i(0) = 0, & \forall i \in I_0(\prtition'), \label{eq:origin lyap} \\
    V_i(v_k) = V_j(v_k), & \forall v_k \in \Vertex(X_i) \cap \Vertex(X_j), i \neq j, \label{eq:cont lyap like} \\
    \tau_i \geq 0, & \forall i \in I(\prtition'), \label{eq:PS lyap}
\end{align}
\end{subequations}
where $\tau_i$ is the slack variable added to soften constraint~\eqref{eq:neg der}. Constraint~\eqref{eq:neg der} ensures the function decreases along $\PWA$ dynamics~\eqref{eq:pwa dynamic},~\eqref{eq:origin lyap} sets $V(0) = 0$, and~\eqref{eq:cont lyap like} enforces continuity.

The optimization problem~\eqref{eq:Lyap-lik opt} is always feasible due to the slack variables $\tau_i$. However, a valid solution requires $\sum_{i=1}^{N} \tau_i = 0$. If this condition is not met, refinement is applied as described in~\cite{10313502}. If a valid Lyapunov-like function within the certified invariant set is found, then the invariant set can be considered as an estimation of the RoA.

To summarize, SEROAISE used IISE as a core to obtain the invariant subset for a $\PWA$ dynamic and combine it with a $\PWA$ Lyapunov-like function search, validated using~\eqref{eq:Lyap-lik opt}. A detailed implementation is provided in Algorithm~\ref{alg:SEROAISE}.

There are significant advantages to this method over previous approaches~\cite{10313502, 10108069}. One of the advantages of the proposed method is that it enforces the Lyapunov-like condition within the invariant set. Furthermore, the RoA obtained using SEROAISE is significantly larger than that obtained using ~\cite {10313502,10108069}. The IISE algorithm in SEROAISE is directly responsible for this larger RoA. In order to demonstrate the performance of the SEROAISE algorithm, in the following section, non-trivial examples are presented, and their results are compared with the state-of-the-art. 
\begin{algorithm}[tb]
    \begin{algorithmic}
    \REQUIRE  $\PWA(x)$ dynamic~\eqref{eq:pwa dynamic}.
    \STATE{Utilizing IISE as described in section~\ref{subseq:IISE} to find the $\PWA$ certified invariant subset for the $\PWA$ dynamic~\eqref{eq: BF PWA} over $\Domain$.}
    \STATE {Set $\Sigma\tau_i\rightarrow \infty$}
    \STATE {Find the new partition $\prtition'$ as defined in~\eqref{eq:New partition}.}
    \WHILE{$\Sigma \tau_i\neq0$}
    \STATE{Solving optimization problem~\eqref{eq:Lyap-lik opt} to find the Lyapunov-like function on the new partition $\prtition'$.}
    \IF{$\tau_i\neq 0$}
        \STATE{Refine the $\prtition'$ using vector field refinement~\cite{10313502}.}
    \ENDIF
    \ENDWHILE
    \RETURN RoA estimation.
    \end{algorithmic}
    \vspace{1em}
    \caption{\textbf{SEROAISE} Algorithm to find the RoA over a certified invariant set}
    \label{alg:SEROAISE}    
\end{algorithm}
\section{Results and Simulations}
All computational experiments were conducted using Python 3.11 on a system with a 2.1 GHz processor and 8 GB RAM. A tolerance level of $10^{-6}$ was used to classify nonzero values. Additionally, a tolerance of ${\epsilon_1=\epsilon_2=\epsilon_3=10^{-4}}$ was applied in the examples. Optimization was carried out using Gurobi~\cite{gurobi}, and Numba~\cite{lam2015numba} was employed to accelerate the computations.

\begin{exmp}[Path Following~\cite{wu2023neural,yanglyapunov}]\label{ex:path following}
The path following dynamics are modeled as follows:
\begin{align}
    \dot{s} &=  \frac{v \cos(\theta_e)}{1 - \dot{e}_{\text{ra}} \kappa(s)},\nonumber \\
    \dot{e}_{\text{ra}} &= v \sin(\theta_e),\nonumber \\
    \dot{\theta}_e &= \frac{v \tan(\delta)}{L} -  \frac{v \kappa(s) \cos(\theta_e)}{1 - \dot{e}_{\text{ra}} \kappa(s)}.
\end{align} 
where the state variables are $e_{ra}$, distance error, and $\theta_e$, angle error. $\delta$ is the steering angle and $|\delta|\leq 40^\circ$ therefore, we limit the control as $tan(-40^\circ)\leq u\leq tan(40^\circ)$. We used a circular path for this experiment with parameters such as fixed speed $v=2\, \mathrm{m/s} $, curvature $\kappa=0.01$, and length $L=1.0$ as described in~\cite{wu2023neural}.

In this study, the NN-based controller from~\cite{yanglyapunov} is used. System identification for the closed-loop dynamics is performed using a single-hidden-layer neural network with 15 ReLU-activated neurons. The RoA is obtained using \textbf{SEROAISE}, as outlined in \textbf{Algorithm}~\ref{alg:SEROAISE}.
IISE with two iterations is employed in this algorithm to expand the set of invariants obtained by~\cite{samanipour2024invariant}, as illustrated by~\ref{fig:PF_Iter}. Afterward, a Lyapunov-like function is obtained to ensure that the certified invariant set accurately represents RoA estimation.
The results are benchmarked against RoAs obtained using Lyapunov-stable neural networks~\cite{yanglyapunov} and DITL~\cite{wu2023neural}, as shown in Fig.~\ref{fig:comparison_PF}. The results indicate that \textbf{SEROAISE} produces a considerably larger RoA compared to~\cite{yanglyapunov,wu2023neural}.  
The detailed information is summarized in Table~\ref{table:summary}.
\begin{figure*}
     \centering
     \begin{subfigure}[b]{0.48\textwidth}
         \centering
         \includegraphics[width=\textwidth]{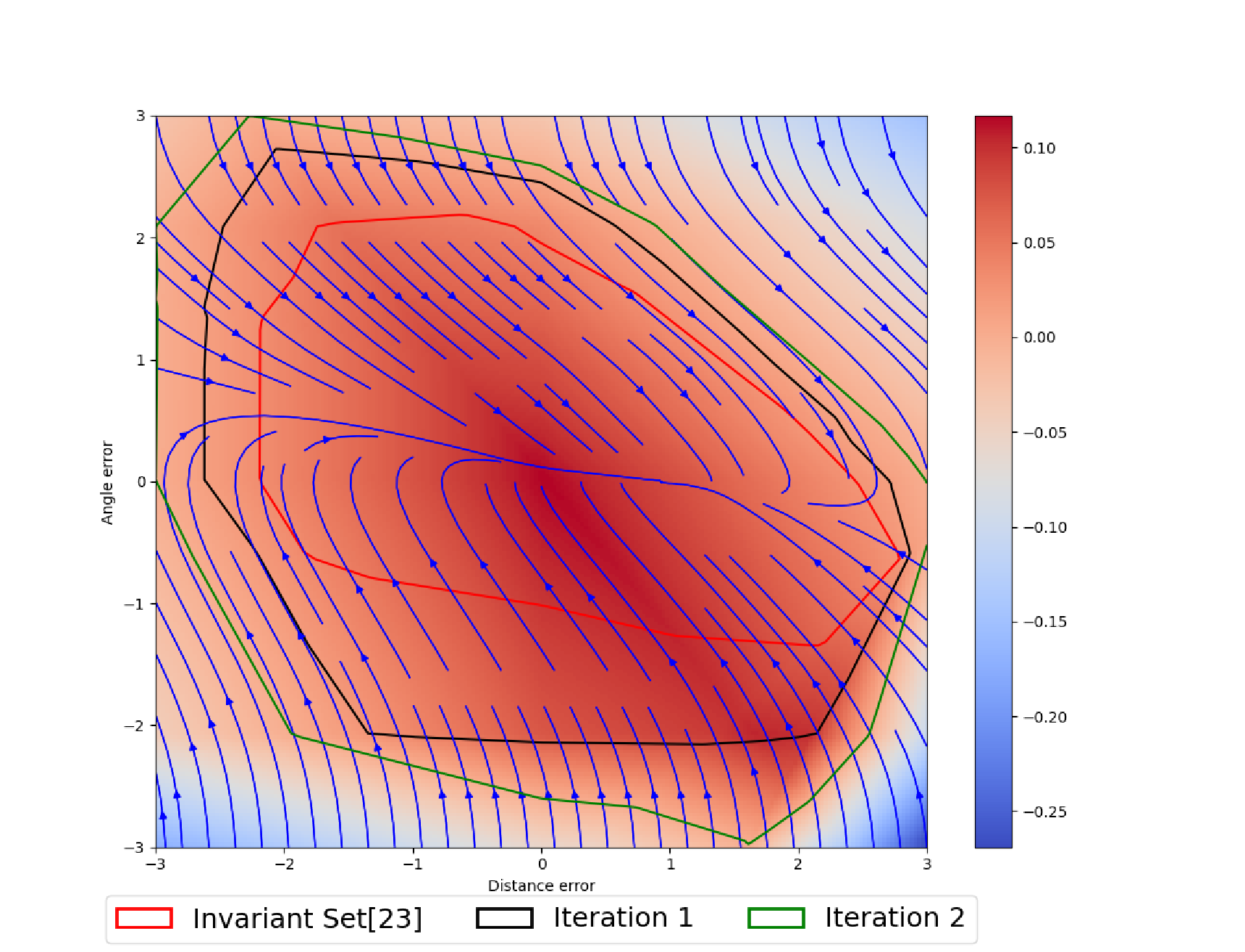}
     \caption{The Certified invariant set obtained for the path-following example~\ref{ex:path following} using the NN controller~\cite{yanglyapunov} after two iterations with \textbf{IISE} Algorithm}
         \label{fig:PF_Iter}
     \end{subfigure}
     \hfill
     \begin{subfigure}[b]{0.48\textwidth}
         \centering
         \includegraphics[width=\textwidth]{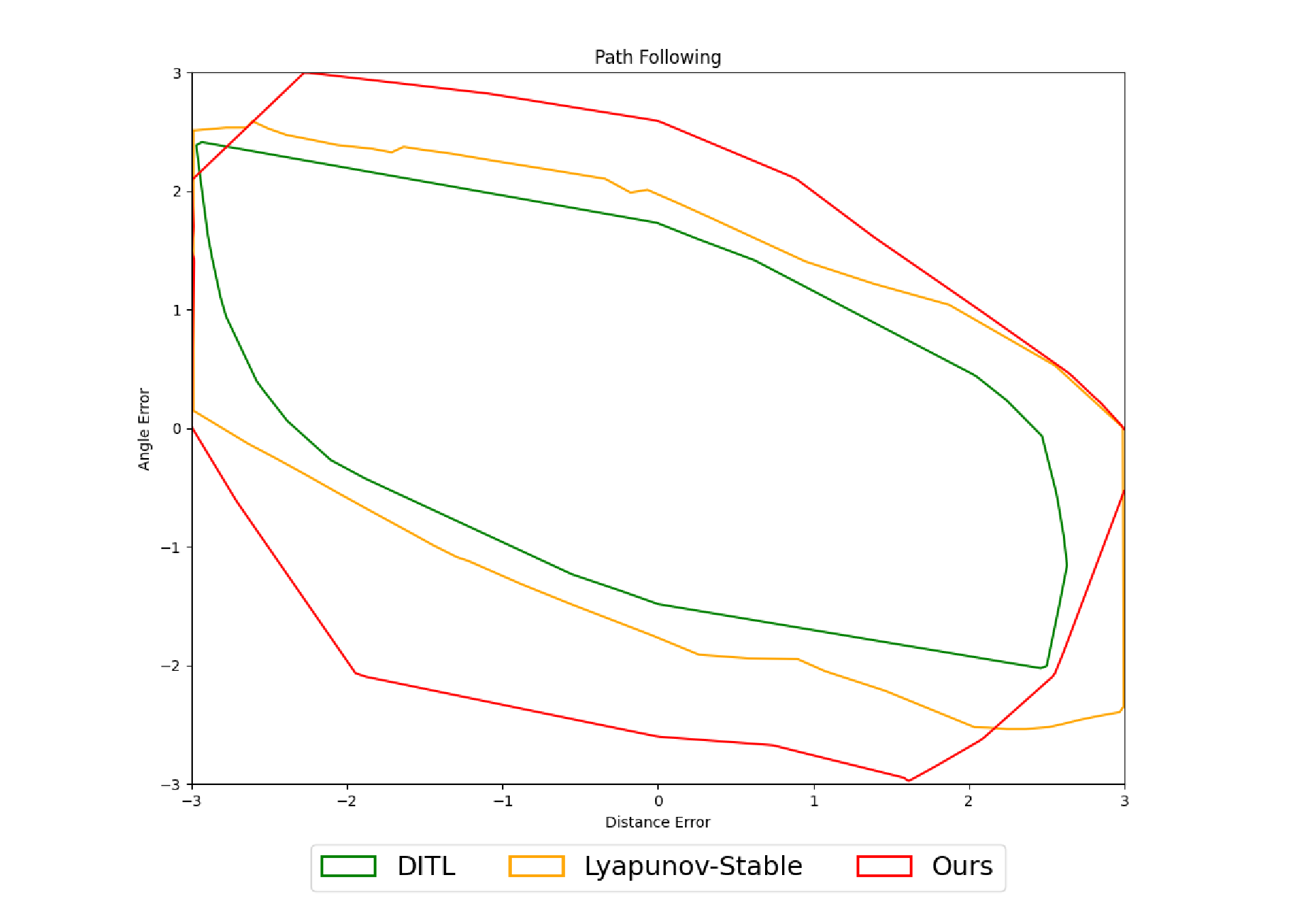}
         \caption{Comparison of the RoA obtained for the path-following using \textbf{SEROAISE} with controller~\cite{yanglyapunov} with methods proposed in~\cite{yanglyapunov} and~\cite{wu2023neural}.}
         \label{fig:comparison_PF}
     \end{subfigure}
    \caption{RoA for the Path-following Example~\ref{ex:path following} determined using SEROAISE Algorithm~\ref{alg:SEROAISE} with NN controller~\cite{yanglyapunov}.}
    \label{fig:comparing PF}
\end{figure*}
\end{exmp}
\begin{exmp}[Inverted Pendulum]\label{ex:IP_new}
We consider the standard dynamics model for an inverted pendulum, as presented in~\cite{chang2019neural, zhou2022neural}:
\begin{equation}
    \ddot{\theta} = \frac{mg\ell \sin(\theta) + u - b\dot{\theta}}{m\ell^2},
\end{equation}
where the physical parameters are defined as follows: $g = 9.81 \, \mathrm{m/s^2}$, $m = 0.15 \, \mathrm{kg}$, $b = 0.1 \, \mathrm{N \cdot m \cdot s/rad}$, and $\ell = 0.5 \, \mathrm{m}$, as described in~\cite{wu2023neural}. The input torque is constrained to $u \in [-6.0, 6.0]$ Nm. 

To stabilize the system, we use an LQR controller with the control law defined as:
\begin{equation}
    u = 1.97725234\theta - 0.97624064\dot{\theta}.
\end{equation}
The closed-loop dynamics with this LQR controller are identified using a single-hidden-layer neural network with 35 ReLU-activated neurons. The proposed \textbf{SEROAISE} algorithm is applied to compute the RoA. In SEROISE, the invariant set obtained by~\cite{samanipour2024invariant} is expanded by IISE in three iterations as shown in Fig.~\ref{fig:Sequential method} and then the Lyapunov-like function is obtained over the certified invariant set. 
Fig.~\ref{fig:comparison IP} illustrates how the obtained RoA can be compared with those obtained using other approaches, such as the Lyapunov-stable approach~\cite{yanglyapunov} and the DITL method~\cite{wu2023neural}. Our results demonstrate that \textbf{SEROAISE} achieves a larger RoA compared to these methods. The total computation time was approximately 150 seconds. A detailed summary is provided in Table~\ref{table:summary}.
\begin{figure*}
    \centering
    \begin{subfigure}[b]{0.49\textwidth}
        \centering
        \includegraphics[width=\textwidth]{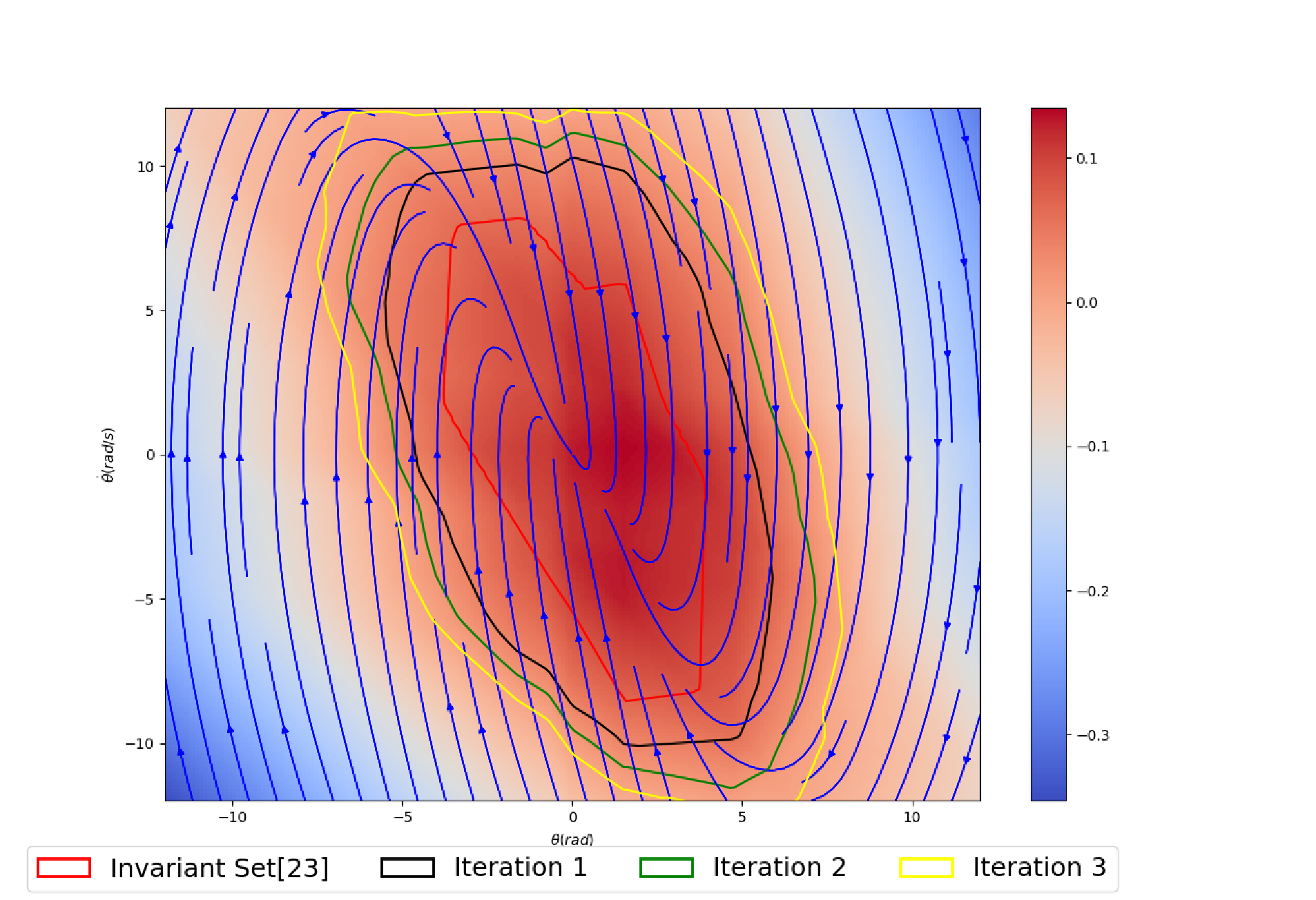}
        \caption{The verified invariant set for the inverted pendulum (Example~\ref{ex:IP_new}) using the LQR controller over three iterations with \textbf{IISE} Algorithm.}
        \label{fig:Sequential method}
    \end{subfigure}
    \hfill
    \begin{subfigure}[b]{0.5\textwidth}
        \centering
        \includegraphics[width=\textwidth]{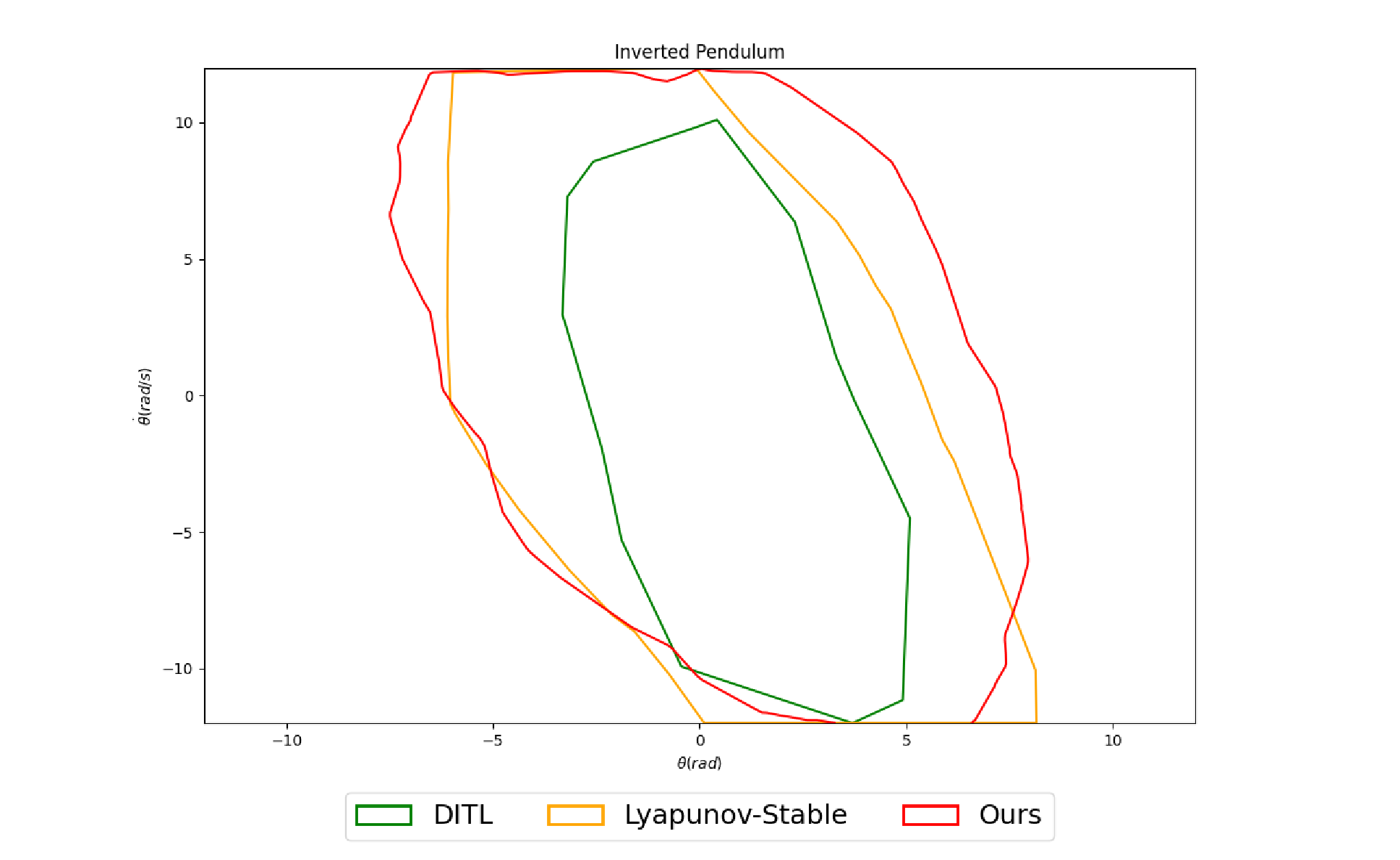}
        \caption{Comparison of the RoA obtained for the inverted pendulum using \textbf{SEROAISE} with LQR controller with methods proposed in~\cite{yanglyapunov} and~\cite{wu2023neural}.}
        \label{fig:comparison IP}
    \end{subfigure}
    \caption{RoA for the inverted pendulum in Example~\ref{ex:IP_new} with LQR controller computed using \textbf{SEROAISE}.}
    \label{fig:comparing IP}
\end{figure*}
\end{exmp}

\begin{exmp}[Cart Pole]\label{ex:Cart pole}
We consider the linearized dynamics of a cart-pole system, controlled by an LQR controller:
\begin{equation}
    \dot{x}(t) = Ax(t) + Bu(t),
\end{equation}
where:
\begin{equation}
A =
\begin{bmatrix}
0 & 1 & 0 & 0 \\
0 & 0 & -0.98 & 0 \\
0 & 0 & 0 & 1 \\
0 & 0 & 10.78 & 0
\end{bmatrix}, \quad
B =
\begin{bmatrix}
0 \\
1 \\
0 \\
-1
\end{bmatrix},
\end{equation}
and $x = [x, \dot{x}, \theta, \dot{\theta}]^T$, where $x$ represents the cart position, and $\theta$ represents the pendulum angle. The LQR control input is given by:
\begin{equation}
    u = x + 2.4109461\dot{x} + 34.36203947\theta + 10.70094483\dot{\theta},
\end{equation}
as described in~\cite{wu2023neural}. The domain of interest is defined as $\mathcal{D} = 1 - |x|_{\infty} \geq 0$ and the control saturation is $-60\leq u\leq 60$.

The closed-loop dynamic is identified using a single-hidden-layer neural network with 15 ReLU-activated neurons. In \textbf{SEROAISE}, the invariant subset is estimated by two iterations, and then the RoA is calculated over the certified invariant set.  
The total computation time for this example was approximately 4178 seconds.  A detailed summary is provided in Table~\ref{table:summary}.
The results highlight the efficiency and scalability of \textbf{SEROAISE} to calculate RoA in complex systems.
\end{exmp}

\def\mtw{\textwidth}
\begin{table}
\vspace{1em}
\centering
\begin{tabular}{p{0.12\mtw}|p{0.08\mtw}|p{0.08\mtw}|p{0.09\mtw}}
\cline{1-4}
Example & Initial $\#$ cells & final $\#$ cells &  Computational time(s)\\ \hline
Path Following~\ref{ex:path following}  & 93 & 162 & 26 \\  \hline
Inverted Pendulum~\ref{ex:IP_new}  & 345 & 2922 & 157 \\ \hline
Cart pole~\ref{ex:Cart pole} &183& 45896& 4178\\ \hline
\end{tabular}	
\caption{Summary of results for finding the certified RoA using \textbf{SEROAISE}~\textbf{Algorithm} as described in~\ref{alg:SEROAISE}.}
\label{table:summary}
\end{table}
\section{Discussion and Limitations}
Our proposed \textbf{SEROAISE} method has demonstrated its versatility by successfully handling a wide range of applications, as evidenced by the example results. In this framework, the RoA is modeled as either a $\PWA$ function or a single-hidden-layer ReLU network—an approach that achieves performance comparable with deeper neural networks while maintaining a relatively straightforward architecture.

Despite these advantages, \textbf{SEROAISE} currently depends on linear programming, which can become computationally intensive for large-scale problems, as outlined in our previous work~\cite{samanipour2024invariant}. This reliance on linear optimization may limit its scalability. To address this challenge, GPU-friendly approaches—such as those discussed in~\cite{yanglyapunov}—could potentially offer computational speedups and improved scalability.

Nevertheless, the novel concept of $\NUGIS$ can be valuable in the case of SEROAISE's scalability problem. Specifically, $\NUGIS$ may be incorporated into other methodologies, such as the approach described in ~\cite{yanglyapunov} to maximize RoA expansion. Specifically, $\NUGIS$ ensures the existence of a larger invariant subset and larger RoA without incurring high computational costs. 
\section{Conclusion}
This paper introduces SEROAISE, an efficient and systematic approach to estimating the Region of Attraction (RoA) in systems described by Piecewise Affine ($\PWA$) functions or equivalent ReLU neural networks. The SEROAISE algorithm consists of two innovative steps: first, it constructs a certified $\PWA$ invariant subset by using the Iterative Invariant Set Estimator (IISE); and second, it conducts a Lyapunov-like search within this subset in order to certify the RoA. SEROAISE achieves both computational efficiency and larger RoAs by avoiding computationally intensive verifiers such as SMT and mixed-integer programming, thereby surpassing conventional methods.

A key element of this work is the concept of Non-Uniform Growth of Invariant Sets ($\NUGIS$), which identifies the potential of non-uniform growth in the invariant set. The efficacy of SEROAISE and $\NUGIS$ is demonstrated through applications including inverted pendulum and path-following systems, where these approaches yield larger ROAs without incurring high computational costs. The framework’s adaptability and reduced reliance on costly verification tools make it highly promising for safety-critical fields such as robotics and autonomous vehicles. Future research may extend the methodology to higher-dimensional problems, or integrate it with advanced control strategies

\begin{ack}                               
This work is supported by the Department of Mechanical Engineering at the University of Kentucky.  
\end{ack}

\bibliographystyle{plain}        
\bibliography{autosam}           



\end{document}